\documentclass[11pt]{article}

\usepackage{amsmath}
\usepackage{amssymb}
\usepackage{appendix}
\usepackage{multirow}
\usepackage{geometry}
\usepackage{float}
\usepackage{xcolor}
\usepackage{cite}
\usepackage{graphicx}
\usepackage{bbm}
\usepackage{enumerate}
\usepackage[numbers]{natbib}
\usepackage{url}
\usepackage[compat=1.1.0]{tikz-feynman}

\newcommand{\Cc}[1]{{\cal C}_{#1}}

\newcommand\CC{{\cal C}}
\def\beq{\begin{equation}}
\def\eeq{\end{equation}}
\def\bea{\begin{eqnarray}}
\def\eea{\end{eqnarray}}
\def\nn{\nonumber}

\def\roughly#1{\mathrel{\raise.3ex\hbox
{$#1$\kern-.75em\lower1ex\hbox{$\sim$}}}}

\def\sla#1{\raise.15ex\hbox{$/$}\kern-.57em #1}

\def\bs{B_s^0}

\def\bsbar{{\bar B}^0_s}

\def\bsmumu{b \to s \mu^+ \mu^-}
\def\bsee{b \to s e^+ e^-}
\def\bstautau{b \to s \tau^+ \tau^-}
\def \NP{{\rm NP}}
\def\bsll{b \to s \ell^+ \ell^-}
\def\bsnunubar{b \to s \nu {\bar\nu}}

\def\bctaunu{b \to c \tau^- {\bar\nu}_\tau}
\def\bclnu{b \to c \ell^- {\bar\nu}_\ell}
\def\bulnu{b \to u \ell^- {\bar\nu}_\ell}

\begin{document}

\begin{flushright}
UdeM-GPP-TH-21-289 \\
\end{flushright}

\begin{center}
\bigskip
{\Large \bf \boldmath $B$ Flavour Anomalies: \\ 2021 Theoretical
  Status Report\footnote{\textmd To be published in the {\it Annual
      Review of Nuclear and Particle Science}}} \\
\bigskip
\bigskip
{\large
David London $^{a,}$\footnote{london@lps.umontreal.ca}}
and Joaquim Matias $^{b,c,}$\footnote{matias@ifae.es} \\
\end{center}

\begin{flushleft}
~~~~~~~~~~~~~~~~$a$: {\it Physique des Particules, Universit\'e de Montr\'eal,}\\
~~~~~~~~~~~~~~~~~~~~~{\it 1375 Avenue Th\'er\`ese-Lavoie-Roux, Montr\'eal, QC, Canada  H2V 0B3} \\
~~~~~~~~~~~~~~~~$b$: {\it Universitat Aut\`onoma de Barcelona, }\\
~~~~~~~~~~~~~~~~~~~~{\it E-08193 Bellaterra (Barcelona), Spain} \\
~~~~~~~~~~~~~~~~$c$: {\it Institut de F{\'\i}sica d'Altes Energies and }\\
~~~~~~~~~~~~~~~~~~~{\it  The Barcelona Institute of Science and Technology, Spain}\\
\end{flushleft}

\begin{center}
\bigskip (\today)
\vskip0.5cm {\Large Abstract\\} 
\vskip3truemm
\parbox[t]{\textwidth}{At the present time, there are discrepancies
  with the predictions of the SM in several observables involving
  $\bsll$ and $\bclnu$ decays.  These are the $B$ flavour
  anomalies. In this review, we summarize the data as of Moriond 2021
  and present theoretical new-physics explanations from both a
  model-independent effective-field-theory point of view and through
  the building of explicit models.  Throughout, we stress the
  complementarity of these two approaches. We also discuss combined
  explanations of both $B$ anomalies, and present models that also
  explain other problems, such as dark matter, $(g-2)_\mu$, neutrino
  properties, and hadronic anomalies.}
\end{center}

\thispagestyle{empty}
\newpage
\setcounter{page}{1}
\baselineskip=14pt

\section{Introduction}

At the present time (2021), there are a number of measurements of
observables involving $B$-meson decays that are in disagreement with
the predictions of the standard model (SM). These are the $B$ flavour
anomalies. There are two types. The neutral-current anomalies are
found in processes involving $\bsll$ decays, while the charged-current
anomalies involve $\bclnu$.

The first observation of the $\bsll$ anomaly was in 2013. At the EPS
conference in Stockholm, the LHCb Collaboration announced
\cite{LHCb:2013ghj} that they had measured the observable $P'_5$
\cite{Descotes-Genon:2012isb} in the four-body angular distribution
$B_d \to K^{*0} (\to K^+\pi^-) \mu^+\mu^-$, and found that it
disagreed with the SM prediction at the level of 3.7$\sigma$
\cite{Descotes-Genon:2013wba}. Soon thereafter, another, smaller
discrepancy of $2.6\sigma$ was found in the measurement of $R_K \equiv
{\cal B}(B^+ \to K^+ \mu^+ \mu^-) / {\cal B}(B^+ \to K^+ e^+ e^-)$
\cite{LHCb:2014vgu}. The SM predicts $R_K \simeq 1$
\cite{Hiller:2003js}, so the experimental result suggests the
violation of lepton flavour universality, i.e., the violation of the
property that the interactions between gauge bosons and charged
leptons are the same for all generations of leptons. Since then, the
number of measured observables involving $\bsll$ decays has grown
enormously, and many of these also exhibit discrepancies with the
SM. What is particularly striking here is the consistency of these
measurements: the discrepancies all go in the same direction, pointing
to a common new-physics explanation.

The first measurement of a $\bclnu$ anomaly was in 2012: the BaBar
Collaboration measured $R_{D^{(*)}} \equiv {\cal B}({\bar B} \to
D^{(*)} \tau^- {\bar\nu}_\tau) / {\cal B}({\bar B} \to D^{(*)} \ell^-
{\bar\nu}_\ell)$ ($\ell=e,\mu$), and found values for $R_D$ and
$R_{D^*}$ that, when taken together, exceed the SM expectation by
$3.4\sigma$ \cite{BaBar:2012obs}. These measurements were repeated by
BaBar (2013) \cite{BaBar:2013mob}, Belle (2015) \cite{Belle:2015qfa}
and LHCb (2015) \cite{LHCb:2015gmp}, and the results were largely
confirmed. Here too, measurements of additional $\bclnu$ observables
have been made subsequently (though far fewer than in the $\bsll$
case), and other deviations from the SM predictions have been found.

Now, the SM has made a great number of predictions, most of which have
been confirmed (e.g., the discovery of the Higgs boson). However, it
also leaves many questions unanswered (what is dark matter?, what is
the origin of the baryon asymmetry of the universe? why are there
three generations?, etc.). For this reason, it is clear that the SM is
not complete -- there must be physics beyond the SM. The $B$ flavour
anomalies have been around for almost ten years. They have been
observed in a variety of processes, and by different experiments. We
may in fact be seeing the first signs of new physics. For this reason,
the study of the $B$ flavour anomalies has become a very exciting
subject. A great deal of work has been done, by many people, with the
aim of finding explanations of the $\bsll$ and $\bclnu$ anomalies,
either individually or simultaneously.

In this review we present a status report on the $B$ flavour anomalies
from a theoretical perspective. For each of the $\bsll$ and $\bclnu$
anomalies, we examine possible solutions from an
effective-field-theory point of view (global analyses), and in the
context of model building. Throughout, we stress the complementarity
of these two approaches. We also review the combined explanations of
the two types of anomalies, as well as models that attempt to also
explain other discrepancies with the SM (dark matter, $(g-2)_\mu$,
etc.). Finally, we take a glance into the future, pointing out
interesting observables whose measurement can distinguish various
explanations.

\medskip

{\bf Notation:} The material in this review is taken from many
different sources, which have different notations. We have tried to
uniformize this notation as follows. In currents, $Q$ and $L$
represent left-handed (LH) $SU(2)_L$ quark and lepton doublets,
respectively, of any generation. Similarly, $u$, $d$ and $e$ represent
right-handed (RH) $SU(2)_L$-singlet up-type quarks, down-type quarks
and charged leptons, respectively, of any generation. For all of these
fields, the chirality is not explicitly indicated. The symbols $b$,
$s$, $c$, $\mu$, $\tau$ and $\nu_\tau$ refer to the actual
particles. If it is necessary to specify the chirality of the $b$, we
write $b_{L,R}$ or $P_{L,R} \, b$, where $P_{L,R}=(1 \mp \gamma_5)/2$
are the LH and RH projection operators, and similarly for the other
particles. Finally, $\ell$ stands for $e$, $\mu$ or $\tau$, unless
otherwise indicated.

\section{EFTs and model building: two complementary approaches}
\label{SMEFTintro}

The $B$ anomalies suggest the presence of physics beyond the SM. The
next question is then: what can this new physics (NP) be? There are
two different ways of addressing this question. The first uses a
model-independent effective-field-theory (EFT) approach: an effective
Hamiltonian is written that contains all dimension-6 operators that
describe the decay of interest ($\bsll$ or $\bclnu$). A global fit to
all the data is performed to find the preferred values of the
coefficients of these operators (the Wilson coefficients (WCs)). The
second approach is to construct models of NP that reproduce the
data. Both methods have their advantages and disadvantages; by
combining the two, one can get a complete idea of possible NP
solutions.

The exchange of NP particles generates new interactions among the SM
particles, in particular the $\bsll$ or $\bclnu$ four-fermion
operators. Now, to date, no new particles have been observed at the
LHC. As a consequence, it is generally believed that the NP, whatever
it is, must be heavy, with masses of ${\cal O}({\rm TeV})$. This is
above the electroweak scale, so that, when the NP particles are
integrated out, the EFT produced is invariant under the full SM gauge
group, $SU(3)_C \times SU(2)_L \times U(1)_Y$. This is the SM
effective field theory (SMEFT). In it, all dimension-4 operators are
those of the SM; higher-dimension operators are due to NP. These
additional operators are suppressed by powers of $\Lambda$, the scale
of NP. It is the SMEFT that provides the connection between the two
approaches.

The operators in the SMEFT are independent. However, a particular
model will generate only a subset of all SMEFT operators, and the
coefficients of these operators will often be related. This means that
a model analysis must take into account additional constraints, not
just those of the $B$ anomalies.

While the SMEFT applies to TeV-scale physics, the EFT used in the
global fits is appropriate for physics at the scale $m_b$. That is, it
is invariant under $SU(3)_C \times U(1)_{em}$ and all particles with
masses greater than $m_b$ have been integrated out. This is the WET
(weak effective theory).

Since they apply to lower-energy physics, all WET operators can be
mapped to SMEFT operators. This has several consequences. First, those
WET dimension-6 operators that also obey $SU(3)_C \times SU(2)_L
\times U(1)_Y$ are related to (combinations of) dimension-6 SMEFT
operators. But those WET operators that do not respect this symmetry
are mapped at tree level to higher-dimensional SMEFT operators,
typically dimension 8.  These dimension-8 operators involve additional
Higgs fields, and the WET operators are generated when the Higgs gets
a vev $v$. Thus, the WET operators are suppressed by an additional
factor of $v^2/\Lambda^2$, and can be neglected.

Another effect, more related to model building, is the following. EFT
analyses often use a basis for the dimension-6 WET operators that is
not the same as that of SMEFT. That is, a dimension-6 WET operator may
be mapped to a linear combination of dimension-6 SMEFT
operators. However, within the SMEFT, these operators are a-priori
independent. Thus, if the EFT analysis favours NP in a WET operator
that is a combination of SMEFT operators, this indicates that there
must be an additional symmetry relating these operators. This is a
clue for the model builders. (And if such a symmetry does not seem to
exist, or is very contrived, then perhaps this EFT scenario will be
disfavoured in spite of the good fit.)

Indeed, using EFT global-fit analyses, a variety of scenarios have
been proposed in which the NP contributes to different WCs. For
several of these scenarios, models can be constructed with the help of
the SMEFT. In addition, by looking at SMEFT operators, one sees that a
simultaneous explanation of both $\bsll$ and $\bclnu$ anomalies is
possible.

The bottom line is that the EFT and model-building approaches are
complementary, and the addition of SMEFT information provides a
complete overview of possible NP explanations of the $B$ flavour
anomalies.

\section{\boldmath Neutral-current anomalies: $\bsll$}

We begin by examining the neutral-current $B$ flavour anomalies
observed in $\bsll$ transitions. In the SM, this decay occurs at the
one-loop level.

\subsection{\boldmath $\bsll$ observables}
\label{observables}

There are two different categories of $\bsll$ observables that exhibit tension with 
the predictions of the SM: 

\begin{enumerate}
  
\item Observables involving only the $b \to s \mu^+\mu^-$
  transition. Here there are two types that differ in their
  sensitivity to hadronic uncertainties and NP:
\begin{itemize}

\item Branching ratios include ${\cal B}({B^0 \to K^{0*}\mu^+\mu^-})$,
  ${\cal B}({B^+ \to K^{*+}\mu^+\mu^-})$, \\ ${\cal B}({B_s \to \phi
  \mu^+\mu^-})$, ${\cal B}({B^+ \to K^+\mu^+\mu^-})$ and ${\cal
  B}({B^0 \to K^{0}\mu^+\mu^-})$.

\item Observables that parametrize the four-body angular distribution
  $B \to K^*(\to K\pi)\mu^+\mu^-$ or $B_s \to \phi(\to K^+K^-)
  \mu^+\mu^-$. These distributions are functions of the three angles
  that characterize the final four-particle state and $q^2$, the
  dilepton invariant mass-squared (see Ref.~\cite{Kruger:2005ep}). The
  $q^2$ spectrum can be split into three regions: (i) large $K^*$
  recoil, $4m_\ell^2 \leq q^2 \leq 9$ GeV$^2$: here, $E_{K^*} \gg
  \Lambda_{QCD}$, this includes the region near the photon pole, (ii)
  charmonium region, $9 < q^2 < 14$ GeV$^2$: this includes the
  charmonium resonances ($J/\psi$, etc.), (iii) low $K^*$ recoil, $14
  < q^2 < (m_B-m_{K^*})^2$ GeV$^2$: here, $E_{K^*} \simeq
  \Lambda_{QCD}$. Each region requires a different treatment of
  hadronic uncertainties.
 
 The angular observables can be further separated into two categories,
 differentiated by their sensitivity to hadronic effects. The
 non-optimized observables (the $S_i$ \cite{Altmannshofer:2008dz}) are
 more sensitive to hadronic uncertainties, while the optimized
 observables (the $P_i$ \cite{Kruger:2005ep,Matias:2012xw,
   Descotes-Genon:2013vna})\footnote{The first optimized observable,
   $A_T^i$, was introduced in Ref.~\cite{Kruger:2005ep}, followed some
   years later by two additional transverse asymmetries in
   Ref.~\cite{Becirevic:2011bp}. A complete basis to describe the
   four-body angular distribution was introduced afterwards in
   Ref.~\cite{Matias:2012xw}, and it was redefined in
   Ref.~\cite{Descotes-Genon:2013vna} to adapt more easily to the
   experimental measurements. A generalization of the
   angular-distribution formalism can be found in
   Ref.~\cite{Gratrex:2015hna}.} are constructed so that the
 dependence on soft form factors cancels exactly at leading
 order. They are usually measured in $q^2$ bins.

\end{itemize}

\item Observables that measure lepton-flavour-universality violation
  (LFUV). The gauge bosons of the SM couple identically (i.e.,
  universally) to charged leptons of different generations. The LFUV
  observables include
\bea
&    R_K = \frac{\displaystyle {\cal B}(B^+ \to K^+ \mu^+ \mu^-)}
    {\displaystyle {\cal B}(B^+ \to K^+ e^+ e^-)} ~~,~~~~ 
    R_{K^*} = \frac{\displaystyle {\cal B}(B \to K^{*} \mu^+ \mu^-)}
    {\displaystyle {\cal B}(B \to K^{*} e^+ e^-)} ~~, & \nn\\ 
    &    R_{\phi} = \frac{\displaystyle {\cal B}(B_s \to \phi \, \mu^+ \mu^-)}
    {\displaystyle {\cal B}(B_s \to \phi \, e^+ e^-)} ~. &
\eea
The SM predicts that all of these ratios equal 1 (up to tiny lepton
mass and electromagnetic effects, see Refs.~\cite{Bordone:2016gaq,
  Isidori:2020acz}). A deviation from this prediction in the
measurement of any of these observables would signal the breaking of
lepton-flavour universality. $R_K$ and $R_{K^*}$ have been measured by
LHCb \cite{LHCb:2021trn,LHCb:2017avl} and Belle
\cite{BELLE:2019xld,Belle:2019oag}. Additional LFUV observables can be
constructed using optimized observables \cite{Capdevila:2016ivx}:
\begin{equation}
  \quad Q_i=P_{i}^{\prime \mu}-P_{i}^{\prime e}
\end{equation}
The SM predicts them to vanish to high accuracy. Two of these -- $Q_{4,5}$ -- have already been measured by Belle \cite{Belle:2016fev}.

\end{enumerate} 

Three other types of observables are usually included in the global fits. (i) There are inclusive decays such as $B \to X_s\mu^+\mu^-$ (these measurements are still not very precise). (ii) ${\cal B}(B_s \to \mu^+ \mu^-)$ has been measured by LHCb, ATLAS and CMS. This observable is particularly interesting due to its reduced hadronic sensitivity and its dependence on a reduced subset of Wilson coefficients (see next section). The discrepancy of this measurement with the SM is at the level of $\sim 2\sigma$. (iii) There are radiative observables such as $B \to X_s \gamma$, $B_s \to \phi \gamma$ and $B \to K^*\gamma$. Further details on the experimental status of all of these observables can be found in Ref.~\cite{experimentalreview}.

\subsection{EFT analysis}

\subsubsection{Operators}

In the most general case, $\bsll$ transitions are described by an effective Hamiltonian \cite{Buchalla:1995vs}\footnote{ This operator basis was chosen in Ref.~\cite{Buchalla:1995vs} because it is useful to describe the SM contributions to $\bsll$. Historically, this basis has also been used to describe the NP contributions. However, other choices of basis are equivalent and equally valid.} 
\begin{equation}
{\cal H}_{\rm eff}=- \frac{4G_F}{\sqrt{2}} \, V_{tb}V_{ts}^* \, \frac{e^2}{16 \pi^2} \sum_{i=1}^{ 12} {\cal C}_{i}  {\cal O}_i ~,
\label{heff}
\end{equation}
where $V_{ij}$ are elements of the Cabibbo-Kobayashi-Maskawa (CKM) matrix, and the operators are\footnote{Note that, in the literature, the scalar and pseudoscalar operators are sometimes multiplied by $m_b$.} 
\bea
&& \hskip-2.2truecm
{\mathcal{O}}_7^{(\prime)} = \frac{m_b}{e} \, (\bar{s} \sigma_{\mu \nu} P_{R(L)} b) F^{\mu \nu} ~, \nn\\
{\mathcal{O}}_{9\ell}^{(\prime)} = (\bar{s} \gamma_{\mu} P_{L(R)} b) (\bar{\ell} \gamma^\mu \ell) & ~,~ &
{\mathcal{O}}_{10\ell}^{(\prime)} = (\bar{s} \gamma_{\mu} P_{L(R)} b) (\bar{\ell} \gamma^\mu \gamma_5 \ell) ~, \nn\\
{\mathcal{O}}_{S\ell}^{(\prime)} =  (\bar{s} P_{R(L)} b) (\bar{\ell} \ell) & ~,~ &
{\mathcal{O}}_{P\ell}^{(\prime)} =  (\bar{s} P_{R(L)} b) (\bar{\ell} \gamma_5 \ell) ~, \nn\\
{\mathcal{O}}_{T\ell} =  (\bar{s} \sigma_{\mu \nu} b) (\bar{\ell} \sigma^{\mu \nu} \ell) & ~,~ &
{\mathcal{O}}_{T5\ell} =  (\bar{s} \sigma_{\mu \nu} b) (\bar{\ell} \sigma^{\mu \nu} \gamma_5 \ell) ~. 
\label{bslloperators}
\eea
With ${\cal H}_{\rm eff}$, the short- and long-distance physics are separated: all the information about the heavy degrees of freedom that have been integrated out (both SM and NP particles) is encoded in the Wilson coefficients of the operators, ${\cal C}_{i}$.

It is useful to relate these operators to those of the SMEFT. To this end, we take linear combinations of them to form operators involving only LH and RH fermions:
\bea
& {\mathcal{O}}_{V\ell}^{ij} = (\bar{s} \gamma_{\mu} P_i b) (\bar{\ell} \gamma^\mu P_j \ell) ~~,~~~~ 
{\mathcal{O}}_{S\ell}^{ij} = (\bar{s} P_i b) (\bar{\ell} P_j \ell) ~~, & \nn\\ 
& {\mathcal{O}}_{T\ell}^i = (\bar{s} \sigma_{\mu \nu} P_i b) (\bar{\ell} \sigma^{\mu \nu} P_i \ell)
~~,~~~~ i,j = L, R ~. & 
\eea
Of these ten WET operators, four of them --
${\mathcal{O}}_{S\ell}^{LL}$, ${\mathcal{O}}_{S\ell}^{RR}$,
${\mathcal{O}}_{T\ell}^L$ and ${\mathcal{O}}_{T\ell}^R$ -- are not
generated at dimension 6 in SMEFT \cite{Alonso:2014csa}; in fact, they
are generated at dimension 8 \cite{Burgess:2021ylu}. They are
therefore suppressed by $v^2/\Lambda^2$ compared to the other six
operators, and can be neglected. The two remaining scalar operators
are often written in the literature as
\beq
{\mathcal{O}}_{S\ell}^{(\prime)} = (\bar{s} P_{R(L)} b)(\bar{\ell} \ell) ~~,~~~~
{\mathcal{O}}_{P\ell}^{(\prime)} = (\bar{s} P_{R(L)} b)(\bar{\ell} \gamma_5 \ell) ~, 
\label{bslloperatorsSP}
\eeq
with conditions imposed on the WCs: $C_{S\ell} = -C_{P\ell}$ and $C'_{S\ell} = C'_{P\ell}$ \cite{Alonso:2014csa}. 

${\cal H}_{\rm eff}$ therefore contains eight operators describing $\bsll$ transitions. The SM contributes mainly to the WCs of three of these: ${\mathcal{O}}_7$, ${\mathcal{O}}_{9\ell}$ and ${\mathcal{O}}_{10\ell}$ (there is also a tiny contribution to $\CC'_7$). The values of their coefficients at the scale $\mu=4.8$ GeV are \cite{Huber:2005ig, Gambino:2003zm, Bobeth:2003at, Misiak:2006ab, Huber:2007vv}
\begin{equation}
    \CC_7^{\rm SM}(\mu_b)=-0.29 ~~,~~~~ \CC_9^{\rm eff SM}(\mu_b)=4.1
    ~~,~~~~ \CC_{10}^{\rm  SM}(\mu_b)=-4.3 ~.
\end{equation}
NP can contribute to all eight operators. The WCs encode the short-distance physics and contain both SM and NP information: $\CC_i = \CC_i^{\rm SM} + \CC_i^{\rm NP}$. All of these operators (or subsets of them) enter the observables discussed above. For instance, the angular and LFUV observables receive contributions from all operators (or combinations of them), radiative observables probe the electromagnetic operators ${\mathcal{O}}^{(\prime)}_{7}$, and ${\cal B}_{B_s \to \mu^+\mu^-}$ is sensitive  to the scalar and vector-axial operators. 

\subsubsection{Methodology}

In Sec.~\ref{observables}, we presented the list of $\bsll$
observables, some of which exhibit deviations from the predictions of
the SM. In order to determine whether these form a coherent set of
deviations that have a common explanation, it is necessary to perform
a global fit of all observables. There are two approaches to
performing such a statistical analysis, the frequentist and the
Bayesian methods. The frequentist approach is most commonly used to
analyze the $\bsll$ anomalies, and this is what we focus on in this
review. (See Refs.~\cite{Kowalska:2019ley, Blake:2019guk,
  Ciuchini:2017mik} for Bayesian analyses.)

Global fits have been performed by a number of groups, but there are
three in particular that have provided regular updates\footnote{Other
  recent global fits can be found in Refs.~\cite{Geng:2021nhg,
    Cornella:2021sby}.}. We refer to these groups as ACDMN
(Alguer\'o/Capde-vila/Descotes-Genon/Matias/Novoa-Brunet)
\cite{Descotes-Genon:2015uva, Alguero:2019ptt, Alguero:2021anc}, AS
(Altmannshofer/Stangl) \cite{Altmannshofer:2021qrr} and HMMN
(Hurth/Mahmoudi/Martinez Santos/Neshatpour)
\cite{Hurth:2021nsi}. (These names appear on the latest update; for
all three groups, other authors have been involved over the years.) In
this review, we compare the methodologies and results of all three
groups, but we occasionally focus a bit more on the results of
ACDMN. In performing a fit, it is necessary to choose a basis for the
$\bsll$ operators, and the basis of Eq.~(\ref{bslloperators}) is
usually used. (However, when the results are compared with the SMEFT,
it will be useful to take linear combinations of these operators.)

The analysis of ACDMN is done using a likelihood approach to fit the deviations from the SM values of the relevant WCs, taking into account the experimental and theoretical uncertainties as well as their correlations  in Gaussian approximation. The  main differences in  the set-ups of the global fits of ACDMN, AS and HMMN are: 
\begin{itemize} 

\item Number of observables. There are two types of fits: (i) a complete fit, including all observables, and (ii) a fit including only the LFUV observables. For ACDMN, the complete fit includes all available data (with one important exception, discussed below), for a total of 246 observables (from LHCb, ATLAS, CMS and Belle). The LFUV fit includes $R_{K,K^*}, Q_{4,5}$ and the radiative decays. In contrast, for their complete fit, AS uses only a subset of the available data, with a total of 130 observables~\cite{talk}. But for their LFUV fit, they use the same observables as ACDMN. Finally, the complete fit of HMMN is reasonably exhaustive, including 173 observables. But the $Q_i$ are left out of the LFUV fit in HMMN.

\smallskip

The one process whose observables are excluded from the analysis of ACDMN, but is included in that of AS and HMMN, is the baryonic decay $\Lambda_b \to \Lambda \mu^+\mu^-$. The reason for its exclusion is as follows. The normalization of ${\cal B}(\Lambda_b \to \Lambda \mu^+\mu^-)$ was taken by LHCb from LEP and Tevatron, and required a correct scaling of the ratio of productions between LHC, LEP and Tevatron. However, while the branching ratio is provided in bins at LHCb, it was never measured in bins at Tevatron. Furthermore, combining LEP and Tevatron data is delicate, given the strong dependence on the $b$-quark production process. Thus, the ratio of productions depends also on the kinematics, so that taking this input from LEP and Tevatron introduces some uncertainty and model dependence in the normalization \cite{Blake:2019guk, LHCb:2015tgy}.

\item Wilson coefficients included in the global fits. The $\bsll$ operators that can receive NP contributions can be separated into three types: electromagnetic (${\mathcal{O}}_7^{(\prime)}$, Eq.~(\ref{bslloperators})), vector/axial vector (${\mathcal{O}}_{9\ell}^{(\prime)}$ and ${\mathcal{O}}_{10\ell}^{(\prime)}$, Eq.~(\ref{bslloperators})), and scalar/pseudoscalar (${\mathcal{O}}_{S\ell}^{(\prime)}$ and ${\mathcal{O}}_{P\ell}^{(\prime)}$, Eq.~(\ref{bslloperatorsSP})). In ACDMN's analysis, all electromagnetic and vector/axial vector operators are included, but not scalar/pseudoscalar operators. In the analysis of AS, the vector/axial vector and scalar/pseudoscalar operators are included, but not electromagnetic operators. And HMMN include all operators.

\item Treatment of the covariance matrix. ACDMN use the covariance matrix in the SM after checking in different points of the NP parameter space that the variation of the covariance matrix is not significant. On the other hand, AS compute a covariance matrix at each point. This is exact for branching ratios which are second-order polynomials in the WCs, but for the angular observables and LFUV ratios it is done approximately by expanding the numerator and denominator up to second-order polynomials, assuming small NP contributions to the WCs compared to the SM. The range of validity of this approximation is roughly the same as taking the SM covariance matrix everywhere.

\end{itemize}

\subsubsection{Hadronic uncertainties}
\label{hadronic}

The impact of hadronic uncertainties on the observables of the $B$ anomalies has been object of very intense debates in the past (for a complete discussion, see Refs.~\cite{Descotes-Genon:2014uoa,Capdevila:2017ert}). However, important progress on these uncertainties from explicit computations has been made recently, and this has clarified the situation. 

There are two main sources of hadronic uncertainties affecting exclusive $B$ decays. The first is form factors (FFs). These are local contributions that parametrize the transition of a $B$ meson to a final-state meson $M$. In our case, we are interested in the FFs for $B \to K^{(*)}$ and $B_s \to \phi$. The second is non-local contributions associated with charm-quark loops. In what follows, we denote these as $T_\mu$. 

The  amplitude for the decay $B\to M\ell^+\ell^-$ has the following structure in the SM: 
\begin{equation}
A(B\to M\ell^+\ell^-)=\frac{G_F \alpha}{\sqrt{2}\pi} V_{tb}V_{ts}^* [({A_\mu}+{ T_\mu}) \bar{u}_\ell \gamma^\mu v_\ell
 +{B_\mu} \bar{u}_\ell \gamma^\mu \gamma_5 v_\ell] ~,
\end{equation}
where ${A_\mu} = -(2m_bq^\nu/{q^2}) \Cc{7} \langle M | \bar{s}\sigma_{\mu\nu}P_R b|B\rangle
 +\Cc{9} \langle M | \bar{s}\gamma_\mu P_L b|B\rangle $ and $
{B_\mu} = \Cc{10} \langle M | \bar{s}\gamma_\mu P_L b|B\rangle$.
%
%
The local contributions from FFs are included in $A_\mu$ and $B_\mu$. At low $q^2$,  these FFs are computed using light-cone sum rules (LCSR) with $B$-meson \cite{Khodjamirian:2010vf} or light-meson distribution amplitudes (DAs) \cite{Bharucha:2015bzk}. Recently, important updates that include higher-twist effects in the $B$-meson DA have been presented in the literature \cite{Gubernari:2018wyi}.  At high $q^2$, FFs are taken mainly from lattice computations. In addition, it is customary to extrapolate the lattice computations to low $q^2$ in order to reduce the uncertainty of the LCSR low-$q^2$ computation. The analysis of ACDMN is done using the $B$-meson DA, while AS and HMMN take the light-meson DA results from Ref.~\cite{Bharucha:2015bzk}.

Turning to the non-local charm-quark loop contributions, the fact that $T_\mu$ is inserted at the same place in the amplitude as $A_\mu$ is problematic: how can we disentangle a possible charm-quark loop contribution from a NP contribution to $\Cc{7}$ or $\Cc{9}^{\rm eff}$? Over the years, this was the subject of intense (often acrimonious) debate.  These non-local charm-quark contributions are particularly important in the region of charmonium resonances between the low- and large-$q^2$ regions, and may in principle affect observables such as branching ratios, $S_i$, $R_{K^*}$ in the presence of NP, or $P_5^\prime$. A number of different approaches were used to tackle this problem. (i) The exchange of one soft gluon was computed within LCSR \cite{Khodjamirian:2010vf}. It was found that this contribution {\it increased} the size of the anomaly in $P_5^\prime$. (ii) Later on, a detailed next-to-leading-order computation was done in Ref.~\cite{Gubernari:2020eft}, but it was found that the correction to this leading-order result is very small. (iii) It is difficult to reliably compute the phase difference between the long-distance charm contribution and the short-distance physics. With this in mind, Ref.~\cite{Capdevila:2017ert} conservatively introduced three nuisance parameters (one for each amplitude), to allow for  constructive and destructive interference between the short-distance and long-distance charm contributions computed in Ref.~\cite{Khodjamirian:2010vf}. This effectively increases the impact of the charm-quark loop contribution, reducing the anomaly. (iv) Another approach was to perform a fit to the resonances (in modulo and phase) and see if the tail of some resonance could explain the deviation in the anomaly bins \cite{Blake:2017fyh}. (v) Finally, Ref.~\cite{Bobeth:2017vxj} adopted an approach based on a dispersive representation  using the data on $J/\psi$ and $\psi(2S)$ to determine the analytic structure of the correlation function of interest and its $q^2$ dependence (up to a polynomial).
In all cases, it was found that it is not possible to explain the $P_5^\prime$ anomaly with $T_\mu$ in the SM. 

In Ref.~\cite{Capdevila:2017ert}, improved QCD factorization (QCDf) is used to include all of the above corrections. The amplitude is decomposed as \cite{Beneke:2001at}
\begin{equation} 
\label{amp}
\langle \ell^+\ell^- {\bar K}^*_i | H_{\rm eff} | \bar{B} \rangle= \sum_{a,\pm} C_{i,a} \xi_{a} + \Phi_{B,\pm} \otimes T_{i,a,\pm} \otimes \Phi_{K^*,a} + {\cal O}(\Lambda_{\rm QCD}/m_B) ~.
\end{equation}
Here, $C_{i,a}$ and $T_{i,a}$ ($a=\perp,\|$) are perturbatively-computable contributions to the various $K^*$ polarizations ($i=0,\perp,\|$) and $\Phi_{B,K^*}$ stands for the light-cone DA of the $B$ and $K^*$ mesons. There are two types of corrections. First, there are the factorizable corrections that can be absorbed into the FFs. These include ${\cal O}(\alpha_s)$ and ${\cal O}(\Lambda_{\rm QCD}/m_B)$ corrections.
Second, there are 
nonfactorizable corrections that cannot be absorbed into the FFs. These also come in two types. There are those that originate from hard gluon exchange; these are calculable within QCDf (including insertions of the operators ${\mathcal{O}}_{1-6}$ and ${\mathcal{O}}_8$ \cite{Beneke:2001at}). And there are those that are power corrections of ${\cal O}(\Lambda_{\textrm{QCD}}/m_B)$, which include the $c\bar{c}$ loops mentioned above. In Ref.~\cite{Capdevila:2017ert}, they are included using the computation of \cite{Khodjamirian:2010vf}, but allowing for a constructive/destructive interference with the short-distance effect. 

 Finally, it is often stated that ``LFUV observables are always clean.'' However, this is only true in the SM because of certain cancellations. In the presence of NP, these cancellations do not occur. This is particularly visible in the case of $R_{K^*}$ (see Fig.~3 in Ref.~\cite{Alguero:2021anc}). Thus, when considering NP, it is important to take into account the hadronic uncertainties (form factors and charm loops). 

To recap, there are two sources of theoretical error in the EFT analysis of the $B$ anomalies: form factors and the non-local charm-loop contributions. There has been great progress in the computations of FFs, using different approaches. In addition, the charm-loop contributions have been studied in full detail using different explicit computational methods. The theoretical error in the $\bsll$ anomaly is now under better control. 

\subsection{Global fits: results}

When a global fit is performed, a hypothesis is made regarding which NP Wilson coefficients are allowed to vary. This fixes the number of  parameters of a fit, $n$, and the number of degrees of freedom $N_{dof}=N_{obs}-n$, where $N_{obs} $ is the number of observables. When the results are obtained, there are two statistical quantities of interest:
\begin{enumerate}[(i)]

\item The p-value of a given hypothesis is a function of $\chi^{2}_{min}$  and $N_{dof}$. This quantity is a measure of the goodness of the fit, i.e., it indicates whether or not the hypothesis provides a good fit. For example, the p-value of the SM fit to the $\bsll$ data is $1.1\%$, which corresponds to a disagreement at the level of $\sim 2.5\sigma$. If the p-value were to fall below $5\sigma$, the SM would be excluded.

\item Pull$_{\rm SM}$ provides very precise information on the comparison of the SM and a given NP hypothesis as explanations of the data: it tells us by how many $\sigma$s the SM hypothesis is disfavoured with respect to a given NP hypothesis. It is related to the difference of the $\chi^{2}_{min}$s of the two fits and to the difference in the $N_{dof}$. This allows us to quantitatively compare the Pull$_{\rm SM}$s of hypotheses with different numbers of free parameters.

\end{enumerate}

\subsubsection{1D, 2D and 6D scenarios}
\label{1D2D6Dscenarios}

We begin with global fits in which one NP WC is allowed to vary. These 1D fits provide useful information: they indicate if
there is a WC (or constrained combination of WCs) in a given basis that is dominant in the explanations of the $B$ anomalies. The three groups have all performed 1D fits, and there are two solutions that are most preferred by all of them: $\Cc{9\mu}^{\rm NP}$ and $\Cc{9\mu}^{\rm NP}=-\Cc{10\mu}^{\rm NP}$\footnote{A third possible solution, $\Cc{10\mu}^{\rm NP}$, gives a very good description of the data in the LFUV fit. However, it fails to provide a good description of the data in the complete fit, particularly when compared with other NP hypotheses.}. The results for the ACDMN, AS and HMMN groups are shown in Tables \ref{tab:results1D_ACDMN}, \ref{tab:results1D_AS} and \ref{tab:results1D_HMMN}, respectively.

Above, we have pointed out the differences among the three groups as regards the methodology (particularly the number of observables included in the fit) and the choice of form factors. Despite these, the results in Tables \ref{tab:results1D_ACDMN}, \ref{tab:results1D_AS} and \ref{tab:results1D_HMMN} are remarkably similar. Two types of fits were done: the complete fit, which includes all observables used by the individual groups, and the LFUV fit, which includes mainly observables sensitive to the breaking of lepton flavour universality. For the LFUV fit, the results are virtually identical. For the complete fit, the results of the ACDMN and HMMN groups are remarkably close. (The excellent agreement between these two different analyses also extends to higher-dimensional fits.) Tiny differences are due to the  small difference in the number of observables (and the inclusion of the baryonic $\Lambda_b$ decay in the HMMN analysis) and the different form factors. Still, the general agreement between these two analyses shows that these effects are rather marginal. On the other hand, there are greater differences with the AS result. AS find that the Pulls for the $\Cc{9\mu}^{\rm NP}$ and $\Cc{9\mu}^{\rm NP}=-\Cc{10\mu}^{\rm NP}$ hypotheses are smaller than for ACDMN and HMMN. Also, they find $\Cc{9\mu}^{\rm NP}=-\Cc{10\mu}^{\rm NP}$ has a marginally larger pull than $\Cc{9\mu}^{\rm NP}$, which is opposite to ACDMN and HMMN. This is almost certainly due to the smaller number of observables in the AS analysis -- particularly the exclusion of the [6,8] bin in all observables\footnote{ In contrast to $\Cc{9\mu}$, the solution $\Cc{9\mu}=-\Cc{10\mu}$ produces a SM-like value for $P_5^\prime$ in both anomaly bins [4,6] and [6,8] (see Fig.~5 in Ref.~\cite{Alguero:2019ptt}). Since the [6,8] bin is problematic for the $\Cc{9\mu}=-\Cc{10\mu}$  solution, and since AS have excluded this bin from their fit, this might explain  the preference of the AS fit for $\Cc{9\mu}=-\Cc{10\mu}$. Note that important progress has been made recently in explicit computations for the whole low-q$^2$ region.} -- and, to a lesser extent, the inclusion of $\Lambda_b \to \Lambda \mu^+\mu^-$ at low-recoil\footnote{This mode shows an unexpected preference at low-recoil for $\Cc{9\mu}^{\rm NP}>0$, in contradiction with the rest of the data. However, a recent reinterpretation excluding LEP data finds a result in nice agreement with the rest of the data: it shows a small deficit in this channel instead of an excess \cite{Blake:2019guk}.}.

\begin{table*}[!ht] \footnotesize
    \centering
\begin{tabular}{c||c|c|c||c|c|c} 
ACDMN & \multicolumn{3}{c||}{Complete fit: 246 observables} &  \multicolumn{3}{c}{LFUV + radiative + $B_s \to\mu^+\mu^-$}\\
\hline
1D Hyp.   & Best fit    & Pull$_{\rm SM}$ ($\sigma$) & p-value & Best fit   & Pull$_{\rm SM}$ ($\sigma$) & p-value\\
\hline\hline
\multirow{1}{*}{$\Cc{9\mu}^{\rm NP}$}    & \multirow{1}{*}{$-1.06^{+0.15}_{-0.14}$
} 
&    \multirow{1}{*}{7.0}   & \multirow{1}{*}{39.5\,\%}
&   \multirow{1}{*}{$-0.82^{+0.22}_{-0.24}$}  
& 
\multirow{1}{*}{4.0}  & \multirow{1}{*}{36.0\,\%}  
\\[1mm]
 \multirow{1}{*}{$\Cc{9\mu}^{\rm NP}=-\Cc{10\mu}^{\rm NP}$}    &   \multirow{1}{*}{$-0.44^{+0.07}_{-0.08}$} &   
 \multirow{1}{*}{6.2}  & \multirow{1}{*}{22.8\,\%}
 &  \multirow{1}{*}{$-0.37^{+0.08}_{-0.09}$}   &  
 \multirow{1}{*}{4.6}   & \multirow{1}{*}{68.0\,\%}  \\
\end{tabular} 
\medskip
\caption{Analysis of the ACDMN group as of Moriond 2021: most preferred 1D patterns of NP in $\bsmumu$. For a complete set of 1D patterns,  see Ref.~\cite{Alguero:2021anc}. The $p$-value of the SM hypothesis is $1.1\%$ (corresponding to { a discrepancy with the data of} 2.5$\sigma$) for the complete fit and $1.4\%$ for the LFUV fit.} 
\label{tab:results1D_ACDMN}
\bigskip
    \centering
\begin{tabular}{c||c|c||c|c} 
AS & \multicolumn{2}{c||}{Complete fit: 130 observables}
 &  \multicolumn{2}{c}{LFUV + $B_s \to \mu^+\mu^-$}\\
\hline
1D Hyp.   & Best fit    & Pull$_{\rm SM}$ ($\sigma$)  & Best fit   & Pull$_{\rm SM}$ ($\sigma$) \\
\hline\hline
\multirow{1}{*}{$\Cc{9\mu}^{\rm NP}$}    & \multirow{1}{*}{$-0.80\pm 0.14$} 
&    \multirow{1}{*}{5.7}  
&   \multirow{1}{*}{$-0.74^{+0.20}_{-0.21}$}  
& 
\multirow{1}{*}{4.1}  
\\
 \multirow{1}{*}{$\Cc{9\mu}^{\rm NP}=-\Cc{10\mu}^{\rm NP}$}    &   \multirow{1}{*}{$-0.41 \pm 0.07$} &   
 \multirow{1}{*}{5.9} 
 &  \multirow{1}{*}{$-0.35 \pm 0.08$}   &  
 \multirow{1}{*}{4.6}  
 \\
\end{tabular} 
\medskip
\caption{Analysis of the AS group as of Moriond 2021: most preferred 1D patterns of NP in $\bsmumu$, see Ref.~\cite{Altmannshofer:2021qrr}, v2 [Apr.2021]. The number of observables is taken from Ref.~\cite{talk}. Note: a revised Ref.~\cite{Altmannshofer:2021qrr}, v3 [Sept.2021] finds very similar results: $\Cc{9\mu}^{\rm NP} = -0.73 \pm 0.15$ (Pull$_{\rm SM} = 5.2\sigma$) and $\Cc{9\mu}^{\rm NP}=-\Cc{10\mu}^{\rm NP} = -0.39 \pm 0.07$ (5.6$\sigma$) for the complete fit, and unchanged results for the LFUV fit.}  
\label{tab:results1D_AS}
\bigskip
    \centering
\begin{tabular}{c||c|c||c|c} 
HMMN & \multicolumn{2}{c||}{Complete fit: 173 observables} &  \multicolumn{2}{c}{only $R_{K^{(*)}}$+ $B_{d,s} \to \mu^+\mu^-$}\\
\hline
1D Hyp.   & Best fit    & Pull$_{\rm SM}$ ($\sigma$)  & Best fit   & Pull$_{\rm SM}$ ($\sigma$) \\
\hline\hline
\multirow{1}{*}{$\Cc{9\mu}^{\rm NP}$}    & \multirow{1}{*}{$-0.95 \pm 0.12$} 
&    \multirow{1}{*}{7.6}  
&   \multirow{1}{*}{$-0.77 \pm 0.21$}  
& 
\multirow{1}{*}{4.0}  
\\
 \multirow{1}{*}{$\Cc{9\mu}^{\rm NP}=-\Cc{10\mu}^{\rm NP}$}    &   \multirow{1}{*}{$-0.49\pm 0.08$} &   
 \multirow{1}{*}{6.7} 
 &  \multirow{1}{*}{$-0.38 \pm 0.09$}   &  
 \multirow{1}{*}{4.6}  
 \\
\end{tabular} 
\medskip
\caption{Analysis of the HMMN group as of Moriond 2021: most preferred 1D patterns of NP in $\bsmumu$. For a complete set of 1D patterns,  see Ref.~\cite{Hurth:2021nsi}.}
\label{tab:results1D_HMMN}
\end{table*}

Although the two most-preferred NP explanations are $\Cc{9\mu}^{\rm NP}$ and $\Cc{9\mu}^{\rm NP}=-\Cc{10\mu}^{\rm NP}$, there are other 1D hypotheses that have significant Pulls. ACDMN find that ${\cal C}_{9\mu}^{\rm NP} =-{\cal C}_{9\mu}^{\rm \prime NP}$ has a large pull  in the complete fit, which is about the same size as that of $\Cc{9\mu}^{\rm NP}$ and $\Cc{9\mu}^{\rm NP}=-\Cc{10\mu}^{\rm NP}$. However, this solution predicts $R_K \simeq 1$, as in the SM, so in the LFUV fit, its pull is 3.0$\sigma$, smaller than that of $\Cc{9\mu}^{\rm NP}$ and $\Cc{9\mu}^{\rm NP}=-\Cc{10\mu}^{\rm NP}$. For this reason, this solution is not favoured.

It is interesting to see how the Pulls of the 1D explanations to the $\bsll$ anomalies have evolved in the past eight years, as more observables, more experiments and  better experimental and theoretical precision are included in each iteration. (Here we refer to Refs.~\cite{Descotes-Genon:2015uva,Capdevila:2017bsm,Alguero:2019ptt,
  Alguero:2021anc}.) For the complete fit, the Pull$_{\rm SM}$ of $\Cc{9\mu}^{\rm NP}$ has been 4.5 [2016], 5.8 [2018], 5.6 [2019], 6.3 [2020], 7.0 [2021]. And for $\Cc{9\mu}^{\rm NP}=-\Cc{10\mu}^{\rm NP}$: 4.2 [2016], 5.3 [2018], 5.2 [2019], 5.8 [2020], 6.2 [2021]. In both cases, the size of the discrepancy with the SM of the $\bsll$ anomaly has steadily increased. Note that the hierarchy in Pulls between these two solutions is reversed if only the subset of LFUV observables is considered. A solution \cite{Alguero:2018nvb} to this apparent contradiction is discussed in the next subsection. 

2D fits allow us to explore if, by permitting an extra NP Wilson coefficient (or a constrained combination) to vary, we get a better description of the data than with a single coefficient, or the improvement is negligible\footnote{We stress that, because Pull$_{\rm SM}$ includes information about the $N_{dof}$, one can compare fits with different $N_{dof}$.}. For the 2D fits, we focus on the ACDMN analysis, but we note that AS and HMMN also perform 2D fits. 

A number of 2D hypothesis are examined in Ref.~\cite{Alguero:2021anc}. Here we discuss two of them:
\begin{itemize}

\item[a)] Adding ${\cal C}_{10\mu}^{\rm NP}$ to ${\cal C}_{9\mu}^{\rm NP}$ slightly decreases the Pull$_{\rm SM}$ in the complete fit (Table \ref{tab:results2Dproc}). That is, the addition of this NP WC does not provide a better description of the data. Still, it is clear that a small contribution to ${\cal C}_{10\mu}^{\rm NP}$ (or to ${\cal C}_{10\mu}^{\rm \prime NP}$ or a scalar operator) is required to explain the small ($\sim 2\sigma$) deficit with respect to the SM observed  in ${\cal B}({B_s \to\mu^+\mu^-})$. On the other hand, the recent experimental update by LHCb is closer to the SM value, and this has slightly reduced its significance: in the 2D $[{\cal C}_{9\mu}^{\rm NP},{\cal C}_{10\mu}^{\rm NP}]$ plot, ${\cal C}_{10\mu}^{\rm NP}$ is now consistent with zero at 1$\sigma$.  

\item[b)] If instead a RH current with the structure $\CC_{9\mu}^{\rm \prime NP} = -\CC_{10\mu}^{\rm \prime NP}$ is added to ${\cal C}_{9\mu}$ (hypothesis 5 in notation of \cite{Alguero:2019ptt}), a better description of the data is found: this is the scenario with the largest Pull$_{\rm SM}$ ($7.4\sigma$), see Table \ref{tab:results2Dproc}. The same result is obtained if only $\CC_{10\mu}^{\rm \prime NP}$ is added. The reason is that the RH current counterbalances the effect of a too-large $\CC_{9\mu}^{\rm NP}$ contribution in $R_K$. 
On the other hand, if this RH current structure is added to the hypothesis $\CC_{9\mu}^{\rm NP} = -\CC_{10\mu}^{\rm NP}$ instead of $\CC_{9\mu}^{\rm NP}$, the description of the data is significantly worse with a reduction of nearly 1.5~$\sigma$ in Pull$_{\rm SM}$ (see hypothesis 4 in Ref.~\cite{Alguero:2021anc}). 

\end{itemize}

\begin{table*}[!ht] 
    \centering
\footnotesize
\begin{tabular}{c||c|c|c||c|c|c} 
ACDMN & \multicolumn{3}{c||}{Complete fit: 246 observables} &  \multicolumn{3}{c}{LFUV + radiative}\\
\hline
 2D Hyp.  & Best fit  & Pull$_{\rm SM}$ ($\sigma$)  & p-value & Best fit & Pull$_{\rm SM}$ ($\sigma$) & p-value\\
\hline\hline
$(\Cc{9\mu}^{\rm NP},\Cc{10\mu}^{\rm NP})$ & $(-1.00,+0.11)$ & 6.8 & 39.4\,\% & $(-0.12,+0.54)$ & 4.3 & 65.6\,\% \\
$(\Cc{9\mu}^{\rm NP},\CC_{10\mu}^{\rm \prime NP})$  & $(-1.26,-0.35)$ & 7.4 & 55.9\,\% & $(-1.82,-0.59)$ & 4.7 & 84.1\,\% \\
Hyp.~4 
& $(-0.48,+0.11)$ & 6.0 & 24.0\,\% & $(-0.46,+0.15)$ & 4.5 & 74.5\,\% \\
Hyp.~5 
& $(-1.26,+0.25)$ & 7.4 & 55.8\,\% & $(-2.08,+0.51)$ & 4.7 & 86.0\,\% \\
\end{tabular}
\bigskip
\caption{Analysis of the ACDMN group: most preferred 2D patterns of NP in $\bsmumu$. Hyp.~4 is $(\Cc{9\mu}^{\rm NP}=-\Cc{10\mu}^{\rm NP} , \Cc{9\mu}^{\rm \prime NP} =-\Cc{10\mu}^{\rm \prime NP})$, hyp.~5 is $(\Cc{9\mu}^{\rm NP} , \Cc{9\mu}^{\rm \prime NP} = -\Cc{10\mu}^{\rm \prime NP})$. For a complete set of 2D patterns, see Ref.~\cite{Alguero:2021anc}.} 
\label{tab:results2Dproc}
\end{table*}


ACDMN also performs a 6D fit\footnote{We note that HMMN does 10D and 20D fits \cite{Hurth:2021nsi}.}, see Table \ref{tab:Fit6D}. This shows consistency with the 1D and 2D fits, and confirms that $\Cc{9\mu}^{\rm NP}$ is the dominant NP WC. It is nonzero at more than 3$\sigma$, while all other WCs are consistent with zero at 1$\sigma$. Finally, we note that the Pull$_{\rm SM}$ for this 6D fit has increased in time: 5.1$\sigma$ [2019], 5.8$\sigma$ [2020] to 6.6$\sigma$ in the most recent analysis of 2021.

\begin{table*}[!ht] \scriptsize
{\footnotesize
    \centering
\begin{tabular}{c||c|c|c|c|c|c}
 ACDMN & $\Cc7^{\rm NP}$ & $\Cc{9\mu}^{\rm NP}$ & $\Cc{10\mu}^{\rm NP}$ & $\CC_{7}^{\rm \prime NP}$ & $\CC_{9\mu}^{\rm \prime NP}$ & $\CC_{10\mu}^{\rm \prime NP}$  \\
\hline\hline
Best fit & +0.01 & -1.21 & +0.15 & +0.01 & +0.37 & -0.21 \\ \hline
1$\sigma$ region &\!\! $\!\![-0.02,+0.04]\!\!$ \!\!&\!\! $\!\![-1.38,-1.01]\! $\! & $\!\![+0.00,+0.34]\!\!$ \!\!& \!\! $\!\![-0.02,+0.03]\!\!$ \!\!&\!\! $\![-0.12,+0.80]\!$\! &\!\! $\!\![-0.42,+0.02]\!$ 
\end{tabular}
\medskip
\caption{Analysis of the ACDMN group: best-fit points and 1~$\sigma$ regions for  the 6D hypothesis of NP in $\bsmumu$. The Pull$_{\rm SM}$ of this hypothesis is 6.6$\sigma$ \cite{Alguero:2021anc}.}
\label{tab:Fit6D}
}\end{table*}



\subsubsection{LFUV and LFU New Physics}
\label{LFUVLFU}

As shown in Sec.~\ref{observables}, the $\bsll$ anomalies appear in two types of observables, those that involve only $\bsmumu$, and those that are LFUV. The scenarios of Sec.~\ref{1D2D6Dscenarios} assume that the NP affects only $\bsmumu$. In Ref.~\cite{Alguero:2018nvb} (essentially the ACDMN group) a new NP scenario was explored. Just as there are two categories of observables, it was assumed that there are two types of NP, one LFUV, affecting only $\bsmumu$, the other lepton flavour universal (LFU), equally affecting all $\bsll$ processes, $\ell = e$, $\mu$ and $\tau$. The LFUV NP then explains the LFUV observables, while the combination of LFUV and LFU NP explains the $\bsmumu$ observables.

The WCs of the LFUV and LFU NP are denoted $\CC_{i\mu}^V$ and $\CC_{i}^U$, respectively. The transformation from the standard $(\CC_{i\mu}, \CC_{ie})$ basis to the new  $(\CC_{i\mu}^V, \CC_{i}^U)$ basis is given by 
\begin{eqnarray}
\CC_{i\mu}&=&\CC_{i\mu}^V+\CC_i^U ~, \nn \\
\CC_{i e}&=&\CC_i^U ~,
\end{eqnarray} 
and similarly for $\tau$. In Sec.~\ref{1D2D6Dscenarios}, we saw that the complete fits prefer the NP scenario $\Cc{9\mu}^{\rm NP}$, while the LFUV fits prefer $\Cc{9\mu}^{\rm NP} = - \Cc{10\mu}^{\rm NP}$. This observation suggests looking at the 2D scenario\footnote{For notational convenience,  the subscript $\mu$ of the LFUV contribution is understood, and is not written explicitly.}.
\begin{eqnarray}
[\CC_{9}^V=-\CC_{10}^V,\CC_{9}^U]
\label{LFUVLFUsoln}
\end{eqnarray}
The global fit was performed with this hypothesis~\cite{Alguero:2021anc}, and it was found that it is one of the most promising scenarios, with a Pull$_{\rm SM}$ of 7.3$\sigma$ (scenario 8 in the notation of
Ref.~\cite{Alguero:2018nvb}). As expected, the first component nicely explains the LFUV observables, while the second universal contribution adds a necessary extra contribution to the $b \to s \mu^+\mu^-$ observables (but also $b \to s e^+e^-$). This universal NP contribution helps in explaining all the anomalies (in optimized observables  
for $B\to K^* \mu^+\mu^-$ and $B_s \to \phi\mu^+\mu^-$) that the naive scenario $\CC_{9\mu}^V=-\CC_{10\mu}^V$ cannot explain alone. 
More examples of LFUV-LFU scenarios can be found in Refs.~\cite{Alguero:2018nvb,Alguero:2019ptt,Alguero:2021anc}. 

This idea is not dissimilar from the idea of allowing LFUV NP in $\bsmumu$ and $\bsee$. Complete analyses of this type of explanation, including many different scenarios, can be found in Refs.~\cite{Hurth:2021nsi,Kumar:2019qbv, Datta:2019zca}.

In summary, for the ACDMN group, the title of ``most promising NP scenario'' is no longer a contest between the 1D hypotheses $\Cc{9\mu}^{\rm NP}$ and $\Cc{9\mu}^{\rm NP} = - \Cc{10\mu}^{\rm NP}$, but rather between two 2D hypotheses:
\begin{equation}
\label{finalscenarios}
 \quad \quad   
 [\Cc{9\mu}^{\rm NP} , \Cc{9\mu}^{\rm \prime NP} = -\Cc{10\mu}^{\rm \prime NP}]
 \quad \quad {\rm or} \quad \quad 
 [{\cal C}^V_{9}=-{\cal C}^V_{10}, {\cal C}^U_9]
\end{equation}
Indeed, this confirms what was observed in the 1D fits: a vectorial coupling to leptons in the form of $\Cc{9\mu}^{\rm NP}$ (only for muons) or $\CC^V_{9}$ (universal) is required to have a good description of the data. In this sense, both hypotheses in Eq.~(\ref{finalscenarios}) provide an excellent explanation for the two largest anomalies, $R_K$ and $P_5^\prime$, in contrast to $\Cc{9\mu}^{\rm NP} = - \Cc{10\mu}^{\rm NP}$, which was unable to explain $P_5^\prime$ (see Fig.~4 in Ref.~\cite{Alguero:2021anc}).

\subsection{Scale of New Physics}
\label{ScaleNP}

Flavour observables are sensitive to higher scales than direct searches at colliders. Therefore, if NP affects flavour, it is not surprising that it would be seen first in this sector. This naturally leads to the question: What is the scale of NP required to explain the $\bsll$ anomalies? The answer to this question depends on the assumption we make about the size of the NP couplings and the way NP enters the $\bsll$ amplitude. 

As discussed in Sec.~\ref{SMEFTintro}, the interaction of NP with the SM particles is described by the SMEFT. The dimension-6 four-fermion operators generated when the NP is integrated out are contained in the effective Hamiltonian $H_{eff}^{\rm NP}= \sum \frac{{\cal O}_i}{\Lambda_i^2}$. Comparing this with Eq.~(\ref{heff}) for three different cases, we find the following NP scales:
\begin{itemize}

\item $\bsll$ transitions induced at tree level, assuming ${\cal O}(1)$ couplings: 
\begin{equation}
\Lambda_i^{\rm Tree}= \frac{4 \pi v}{s_w g} \frac{1}{\sqrt{2 |V_{tb} V_{ts}^*|}} \frac{1}{{|C_i^{\rm NP}|}^{1/2}} \sim \frac{35~{\rm TeV}}{{|C_i^{\rm NP}|}^{1/2}} ~,
\end{equation}

\item $\bsll$ transitions induced at loop level, assuming ${\cal O}(1)$ couplings: 
\begin{equation}
\Lambda_i^{\rm Loop}  \sim \frac{35~{\rm TeV}}{4 \pi {|C_i^{\rm NP}|}^{1/2}}\sim  \frac{3~{\rm TeV}}{ {|C_i^{\rm NP}|}^{1/2}} ~,
\end{equation}

\item Minimal Flavour Violation with CKM-SM: There is a further reduction of $1/5$ in the NP scale due to $\sqrt{|V_{tb} V_{ts}^*|} \sim 1/5$.



\end{itemize}
In all cases, if the NP coupling is less than ${\cal O}(1)$, the NP scale $\Lambda$ is reduced correspondingly.


Turning to the $b \to c \tau\nu$ anomalies, this transition takes place at tree level in the SM. As we will see below, an explanation requires a rather large ${\cal O}(10\%)$ tree-level NP contribution at the amplitude level. This implies that
\beq
\Lambda^{\rm NP}\sim 1/(\sqrt{2} G_F |V_{cb}|  0.10)^{1/2} \sim 4~{\rm TeV} ~.
\eeq



\subsection{Mapping WET to SMEFT}

In the fits we have determined which operators must receive NP contributions in order to reproduce the data. There are two types of solutions: (i) pure LFUV, with NP only in $\bsmumu$, and (ii) $\bsmumu$ LFUV + LFU contributions. In both cases, the question now is: what kinds of NP can generate these solutions? In this subsection, we examine general features of such NP -- this is the ``model-independent'' model analysis. Specific models will be examined in the next subsection.

As discussed in Sec.~\ref{SMEFTintro}, the EFT used in the fits is the WET. It is appropriate for physics at the scale $m_b$: it is invariant under $SU(3)_C \times U(1)_{em}$ and all particles with masses greater than $m_b$ have been integrated out. On the other hand, the EFT generated at high energies when the heavy NP particles have been integrated out is the SMEFT. A first step in identifying what kinds of NP can be responsible for the $\bsll$ anomalies comes by mapping WET onto SMEFT, focusing on ${\cal O}^{(\prime)}_{9,10}$.

In the SMEFT, the dimension-6 operators that contribute to these WCs at tree level can be separated into two categories: LFUV and LFU.  The LFUV operators are a subset of the semileptonic four-fermion operators. A simple basis for these operators is \cite{Grzadkowski:2010es}
\bea
{\cal O}_{L Q}^{(1)} &=& ({\bar L}_i \gamma_\mu  L_j) ({\bar Q}_k \gamma^\mu Q_l) ~, \nn\\
{\cal O}_{L Q}^{(3)} &=& ({\bar L}_i \gamma_\mu \sigma^I  L_j) ({\bar Q}_k \gamma^\mu \sigma^I Q_l) ~, \nn\\
{\cal O}_{Q e} &=& ({\bar Q}_i \gamma_\mu Q_j) ({\bar e}_k \gamma^\mu e_l) ~, \nn\\
{\cal O}_{L d}^{(1)} &=& ({\bar L}_i \gamma_\mu  L_j) ({\bar d}_k \gamma^\mu d_l) ~, \nn\\
{\cal O}_{e d}^{(1)} &=& ({\bar e}_i \gamma_\mu e_j) ({\bar d}_k \gamma^\mu d_l) ~,
\label{SMEFTops}
\eea
 where $i,j,k,l$ are flavour (generation) indices. These are LFV for $i\ne j$, but for those that are LFC, one sees that they  can be LFUV since the $ij = ee$, $\mu\mu$ and $\tau\tau$ WCs are not necessarily equal. It is convenient to combine the first two operators: ${\cal O}_{L Q}^{(1,3)} \equiv {\cal O}_{L Q}^{(1)} + {\cal O}_{L Q}^{(3)}$. These four SMEFT operators contribute to the four $\bsmumu$ WCs as follows \cite{Aebischer:2015fzz}:
\bea
\CC_{9\mu}^{\NP} &=& \frac{\pi}{\alpha} \frac{v^2}{\Lambda^2} 
\left[ {\tilde{\CC}}_{L Q}^{(1,3)\mu\mu 23} + {\tilde{\CC}}_{Q e}^{23\mu\mu} \right] ~, \nn\\
\CC_{10\mu}^{\NP} &=& \frac{\pi}{\alpha} \frac{v^2}{\Lambda^2} 
\left[ {\tilde{\CC}}_{Q e}^{23\mu\mu} - {\tilde{\CC}}_{L Q}^{(1,3)\mu\mu 23} \right] ~, \nn\\
\CC_{9\mu}^{\prime \NP} &=& \frac{\pi}{\alpha} \frac{v^2}{\Lambda^2} 
\left[ {\tilde{\CC}}_{L d}^{\mu\mu 23} + {\tilde{\CC}}_{e d}^{\mu\mu 23} \right] ~, \nn\\
\CC_{10\mu}^{\prime \NP} &=& \frac{\pi}{\alpha} \frac{v^2}{\Lambda^2} 
\left[ {\tilde{\CC}}_{e d}^{\mu\mu 23} - {\tilde{\CC}}_{L d}^{\mu\mu 23} \right] ~,
\label{NPops+SMEFT}
\eea
{ where ${\tilde{\CC}}$ indicates a SMEFT WC at low energies, { below the scale of electroweak symmetry breaking}. The ${\tilde{\CC}}$s are related to high-energy SMEFT WCs using renormalization-group running.} { We stress that the above relations hold only for the basis of Eq.~(\ref{SMEFTops}).}

Recall that the preferred 1D fits involve $\CC_{9\mu}^{\NP}$ or $\CC_{9\mu}^{\NP} = -\CC_{10\mu}^{\NP}$ (Tables \ref{tab:results1D_ACDMN}, \ref{tab:results1D_AS} and \ref{tab:results1D_HMMN}). Consider the second solution. This arises if ${\tilde{\CC}}_{Q e}^{23\mu\mu} = 0$. In this case, $\CC_{9\mu}^{\NP} + \CC_{10\mu}^{\NP} = 0$; the only nonzero WC is
\beq
\CC_{9\mu}^{\NP} - \CC_{10\mu}^{\NP} = 2 \frac{\pi}{\alpha} \frac{v^2}{\Lambda^2} 
\, {\tilde{\CC}}_{L Q}^{(1,3)\mu\mu 23} ~.
\eeq
This solution is particularly good for model building. There is only one WC and its presence does not require any special relations between independent SMEFT operators. Also, it has a simple physical interpretation: the NP couples only to LH  fermions.

Compare this with the first solution, $\CC_{9\mu}^{\NP}$. In order for it to be the only nonzero WC, it is necessary that $\CC_{10\mu}^{\NP} = 0$, which requires that
\beq
{\tilde{\CC}}_{L Q}^{(1,3)\mu\mu 23} = {\tilde{\CC}}_{Q e}^{23\mu\mu} ~.
\label{C9condition}
\eeq
Since the operators are independent, their coefficients are generally unrelated; any such relation requires additional explanation, such as a symmetry. 

In the particular case of $\CC_{9\mu}^{\NP}$, such a symmetry is not difficult to find. While the operators ${\cal O}_{L Q}^{(1,3)}$ and ${\cal O}_{Q e}$ both involve LH quarks, they involve LH and RH charged leptons, respectively. For NP in $\bsmumu$ decays, if its couplings are invariant under $L_\mu$ (the $\mu$ lepton number), it will couple equally to $\mu_L$ and $\mu$, and its effects will be seen in  $\CC_{9\mu}^{\NP}$.

{ The point here is that it is possible to find single-particle NP models for each of the 1D solutions of the EFT analysis, $\CC_{9\mu}^{\NP}$ and $\CC_{9\mu}^{\NP} = -\CC_{10\mu}^{\NP}$. Considerations of SMEFT help to accomplish this.}


 For the 2D fits, the preferred scenarios are $(\CC_{9\mu}^{\NP},\CC_{10\mu}^{\prime\NP})$ and $(\Cc{9\mu}^{\rm NP} , \Cc{9\mu}^{\rm \prime NP} = -\Cc{10\mu}^{\rm \prime NP})$ (see Table \ref{tab:results2Dproc}). There is no problem with the second scenario. As discussed above, $\CC_{9\mu}^{\NP}$ can be produced by NP that obeys a symmetry under which $\mu_L$ and $\mu_R$ transform in the same way, and $\Cc{9\mu}^{\rm \prime NP} = -\Cc{10\mu}^{\rm \prime NP}$ is already an SMEFT WC. On the other hand, the first solution is more complicated. First, in order to produce $\CC_{10\mu}^{\prime\NP}$, one requires NP that obeys a symmetry under which $\mu_L$ and $\mu_R$ transform in the {\it opposite} way. That is, two different NP particles are required: the exchange of one generates $\CC_{9\mu}^{\NP}$, and the other generates $\CC_{10\mu}^{\prime\NP}$. Second, it is not obvious what is the symmetry under which $\mu_L$ and $\mu_R$ transform equally but oppositely. 

Finally, the scenario in the 6D fit (Table \ref{tab:Fit6D}) is acceptable from a model-building point of view,  and all WCs can be written in SMEFT form  using the following combinations: $\CC_{9\mu}^{\NP} \pm \CC_{10\mu}^{\NP}$, $\CC_{9\mu}^{\prime\NP} \pm \CC_{10\mu}^{\prime\NP}$, $\CC_{7\mu}^{\NP}$, $\CC_{7\mu}^{\prime\NP}$.

Turning to LFU SMEFT operators that contribute to $\bsll$, there are three dimension-6 operators involving the Higgs field \cite{Grzadkowski:2010es}:
\bea
{\cal O}_{\varphi Q}^{(1)} &=& (\varphi^\dagger i \overleftrightarrow{D}_\mu \varphi ) ({\bar Q}_i \gamma^\mu Q_j) ~, \nn\\
{\cal O}_{\varphi Q}^{(3)} &=& (\varphi^\dagger i \overleftrightarrow{D}_\mu^I \varphi ) ({\bar Q}_i \tau^I \gamma^\mu Q_j) ~, \nn\\
{\cal O}_{\varphi d} &=& (\varphi^\dagger i \overleftrightarrow{D}_\mu \varphi ) ({\bar d}_i \gamma^\mu d_j) ~.
\label{LFUops}
\eea
These generate the following LFU NP contributions \cite{Aebischer:2015fzz}:
\bea 
\CC_{9}^{\NP} &=& \frac{\pi}{\alpha} \frac{v^2}{\Lambda^2} 
\left[ {\tilde{\CC}}_{\varphi Q}^{(1)23} + {\tilde{\CC}}_{\varphi Q}^{(3)23} \right] (-1 + 4 \sin^2 \theta_W ) ~, \nn\\ 
\CC_{10}^{\NP} &=& \frac{\pi}{\alpha} \frac{v^2}{\Lambda^2} 
\left[ {\tilde{\CC}}_{\varphi Q}^{(1)23} + {\tilde{\CC}}_{\varphi Q}^{(3)23} \right] ~, \nn\\ 
\CC_{9}^{\prime \NP} &=& \frac{\pi}{\alpha} \frac{v^2}{\Lambda^2} {\tilde{\CC}}_{\varphi d}^{23} (-1 + 4 \sin^2 \theta_W ) ~, \nn\\ 
\CC_{10}^{\prime\NP}  &=& \frac{\pi}{\alpha} \frac{v^2}{\Lambda^2} {\tilde{\CC}}_{\varphi d}^{23} ~.  
\eea
The preferred LFUV + LFU solution involves $\CC_{9\mu}^{V} = -\CC_{10\mu}^{V}$ and $\CC_{9}^{U}$. From the above, we see that $\CC_{9}^{\NP} \propto \CC_{10}^{\NP}$, so that it is not possible to produce only $\CC_{9}^{U}$ with the operators of Eq.~(\ref{LFUops}). 

\begin{figure}[h]
\begin{center}
\includegraphics[width=0.55\textwidth]{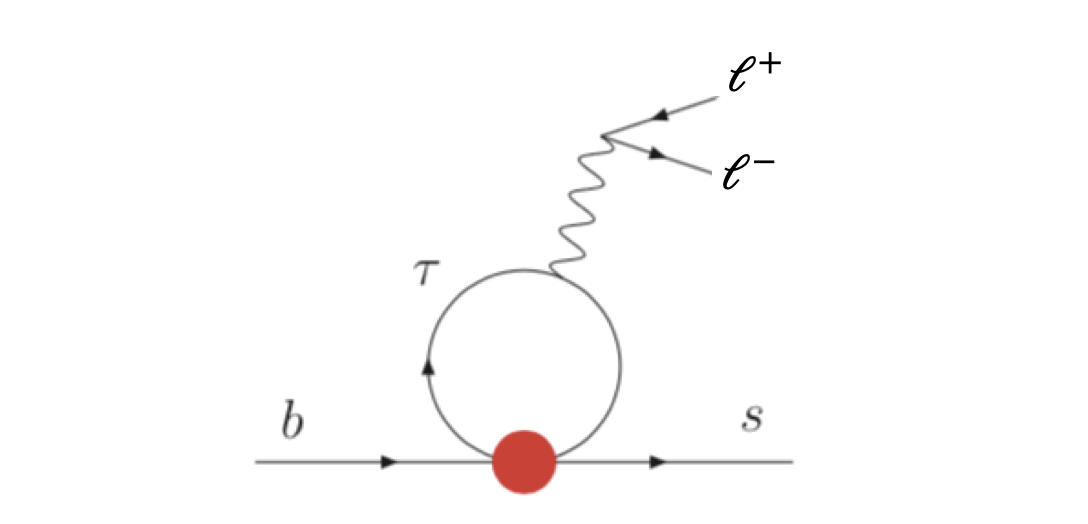}
\end{center}
\caption{Diagram inducing a lepton universal contribution to $\CC_{9}^{U}$ through a $\tau$ loop.}
\label{C9U_loop}
\end{figure}

However, there is another possibility. The full matching of WET to SMEFT must take into account loop effects and renormalization-group running. In particular, a low-energy $\CC_{9}^{U}$ can be generated via a loop-level process  \cite{Crivellin:2018yvo}. The fermion in the loop can be $u_{1,2}$, $d_{1,2,3}$ or $\tau$ \cite{Crivellin:2018yvo, Bobeth:2011st, Alguero:2019ptt, Aebischer:2019mlg}. (For $u_3 = t$, a large ditop-dimuon coupling is disfavoured by EW precision tests \cite{Camargo-Molina:2018cwu}.) For loops with  $u$ or $d$, it must be checked that these are compatible with constraints from dijet searches and non-leptonic decays. On the other hand, the case where the fermion is a $\tau$ (see Fig.~\ref{C9U_loop})  is particularly interesting \cite{Alguero:2019ptt,Capdevila:2017iqn} because it suggests a possible link between NP in the $\bsmumu$ and $\bctaunu$ transitions. The key point here is that, if there is a $\CC_{9\mu}^{V} = -\CC_{10\mu}^{V}$ contribution to $\bsmumu$, it is not difficult to generate $\CC_{9}^{U}$ via a large $b\to s \tau\tau$ effective coupling. Through radiative effects by closing the loop of the operator $b \to s \tau^+\tau^-$, a  NP contribution to $\Cc{9}^{\rm U}$ is induced \cite{Alguero:2019ptt}:
\begin{equation}
    \Cc{9}^{\rm U}\simeq 7.5 \left(1-\sqrt{\frac{R_{D^{(*)}}}{R_{D^{(*) SM}}}}\right) \left(1+ \frac{{\rm log}(\Lambda^2/({\rm 1 TeV^2}))}{10.5}\right)
\end{equation}
where $\Lambda$ is the scale of NP that is taken in a range between 1 to 10 TeV.

\subsection{Models}
\label{bsllmodels}

Many models have been proposed to explain the $\bsmumu$
anomalies. Some are ``simplified models,'' in which NP particles are
added, but without giving any details of the underlying high-energy NP
model from which these particles arise (the UV completion). Others do
present full models, including the full gauge group and its particle
content. In the simplest models (simplified or full), there is a
single NP particle that contributes at tree level to $\bsmumu$. More
complicated models typically involve many NP particles; some of them,
but not all, contribute to $\bsmumu$.

In this review, we describe only the simplest models for all the
anomalies. There are several reasons. First, there is the general
prejudice that Nature prefers simple solutions (although this should
not be taken as a golden rule). But there are also practical
reasons. The complicated NP models have many new particles and
generally have to satisfy many constraints. They are often viable only
in a region of parameter space. But the data have changed over the
years, so it is no longer clear whether these models are still
valid. In contrast, the simplest models have been studied by several
different groups and have been continually updated. (In fact, some of
the early simple models have now been ruled out.) So we are certain
that the models we describe here do actually explain the 2021 data.

For $\bsmumu$, there are two classes of simple models. They involve the tree-level exchange of a $Z'$ or a leptoquark (LQ). These are described below.

\subsubsection{\boldmath $Z'$ models}
\label{Z'models}

In $Z'$ models, the $Z'$ couples to both ${\bar s}b$ and $\mu^+\mu^-$. Given that it
contributes to $\CC_{9\mu}^{\NP}$ or $\CC_{9\mu}^{\NP} = -
\CC_{10\mu}^{\NP}$, the $Z'$ coupling to ${\bar s}b$ is purely LH, but
it may couple to both LH and RH $\mu^+\mu^-$ pairs. 

We write the general couplings of the $Z'$ as
\beq
\left[ g^Q_{ij} ( {\bar Q}_{i} \gamma^\mu Q_{j} )
+ g^{L}_{ij} ( {\bar L}_{i} \gamma^\mu L_{j} )
+ g^{e}_{ij} ( {\bar e}_{i} \gamma^\mu e_{j} ) \right] Z'_\mu ~.
\eeq
For $\bsmumu$, we are interested in those terms whose coefficients in
the mass basis are $g^Q_{23}$, $g^{L}_{22}$ and $g^{e}_{22}$.
When the $Z'$ is integrated out, one generates the following
four-fermion operators relevant for $2q2\ell$ processes (two quarks
and two leptons):
\beq
    {\cal L}^{2q2\ell}_{Z'} = - \frac{g_{23}^Q g^{L}_{22}}{M^2_{Z'}}
    \, ({\bar s}_L \gamma^\mu b_L) \, ({\bar L}_{\mu} \gamma^\mu L_{\mu})
    - \frac{g_{23}^Q g^{e}_{22} }{M^2_{Z'}}
    \, ({\bar s}_L \gamma^\mu b_L) \, ({\bar \mu}_R \gamma^\mu \mu) ~.
\eeq
Both terms contribute to $\bsmumu$, while the first also contributes
to $b \to s {\bar\nu}_\mu \nu_\mu$. If the $\bsmumu$ anomalies are
explained by $\CC_{9\mu}^{\NP}$ alone, then $g^{e}_{22} = g^{L}_{22}$, while $g^{e}_{22} = 0$ if $\CC_{9\mu}^{\NP} = -
\CC_{10\mu}^{\NP}$.

In addition to $2q2\ell$ operators, $Z'$ exchange also produces
four-quark ($4q$) and four-lepton ($4\ell$) operators at tree
level. The operators constructed from the same currents as above are
\bea
    {\cal L}_\NP^{4q} & = &
    - \frac{(g_{23}^Q)^2}{2 M^2_{Z'}} \, ({\bar s}_L \gamma^\mu b_L)\,({\bar s}_L \gamma_\mu b_L) ~, \nn\\
    {\cal L}_\NP^{4\ell} & = & -~\frac{(g_{22}^{L})^2}{2M^2_{Z'}}
    ({\bar L}_{\mu} \gamma^\mu L_{\mu}) ({\bar L}_{\mu} \gamma_\mu L_{\mu})
    -~\frac{g_{22}^{L} g_{22}^{e}}{2M^2_{Z'}}
    ({\bar L}_{\mu} \gamma^\mu L_{\mu}) ({\bar \mu}_R \gamma_\mu \mu) ~.
\eea
The operator in ${\cal L}_\NP^{4q}$ contributes to $\bs$-$\bsbar$
mixing, while those in ${\cal L}_\NP^{4\ell}$ contribute to neutrino
trident production ($\nu_\mu N \to \nu_\mu N \mu^+ \mu^-$)
\cite{Brown:1972vne}.

The point here is that the best-fit values of $\CC_{9\mu}^{\NP}$ or
$\CC_{9\mu}^{\NP} = - \CC_{10\mu}^{\NP}$ place a constraint on
$g_{23}^Q g^{L}_{22} / M^2_{Z'}$. This same quantity is also
constrained by $b \to s {\bar\nu}_\mu \nu_\mu$.\footnote{The decay $b
  \to s {\bar\nu}_\mu \nu_\mu$ is analyzed within the SM in
  Ref.~\cite{Buras:2014fpa}; experimental results are given in
  Ref.~\cite{Belle:2017oht}.}  In addition, there are constraints on
$(g_{23}^Q)^2/M^2_{Z'}$ from $\bs$-$\bsbar$ mixing, and on
$(g_{22}^{L})^2/M^2_{Z'}$ from neutrino trident
production.
  It must be checked that all
these constraints are consistent with one another  (for example, see Ref.~\cite{Alok:2017jgr}). This type of
analysis is necessary for models, but not for EFTs.

Finally, there is an extremely important constraint coming from direct
searches for $Z'$ bosons in the process $p p \to Z' \to \mu^+ \mu^-$
\cite{ATLAS:2019erb}.  The translation of the experimental limit into
a model constraint is completely model-dependent. It depends on how
the $Z'$ couples to all the quarks, and what is the branching ratio of
$Z' \to \mu^+ \mu^-$.  This latter quantity can be affected if the
$Z'$ can also decay into dark matter, for example.

All $Z'$ models must take into account all of these constraints.
(Furthermore, if one is interested in a particular $Z'$ model, it may
be necessary to update the constraint analysis, to see if the model is
still viable.)

Many $Z'$ models have been proposed to explain the $\bsmumu$
anomalies. The following list includes most of them, separated into
four categories: the $Z'$ is heavy and contributes to (i)
$\CC_{9\mu}^{\NP}$, (ii) $\CC_{9\mu}^{\NP} = -\CC_{10\mu}^{\NP}$, or
(iii) $\CC_{9\mu}^{\NP}$ and $\CC_{10\mu}^{\NP}$; (iv) light $Z'$
bosons.
\begin{enumerate}

\item $\CC_{9\mu}^{\NP}$. As argued earlier, this NP explanation can
  arise only if the relation in Eq.~(\ref{C9condition}) holds, which
  generally requires an additional symmetry. $L_\mu$ (the $\mu$ lepton
  number) is such a symmetry, but by itself it is not anomaly-free; an
  additional symmetry is required. The most popular choice is a gauged
  $L_\mu - L_\tau$ symmetry: $Z'$ models with a $U(1)_{L_\mu -
    L_\tau}$ symmetry can be found in
  Refs.~\cite{Altmannshofer:2014cfa, Crivellin:2015mga,
    Crivellin:2015lwa, Altmannshofer:2016jzy, Crivellin:2016ejn,
    Ko:2017yrd, Arcadi:2018tly, Singirala:2018mio, Hutauruk:2019crc,
    Baek:2019qte, Biswas:2019twf, Han:2019diw, Crivellin:2020oup}.
  There are many variants in which the $U(1)$ has a different symmetry, but always includes $L_\mu$, see 
  Refs.~\cite{Ko:2017quv, Bonilla:2017lsq, Bian:2017rpg,
    Duan:2018akc, Geng:2018xzd, Ko:2019tts, Allanach:2020kss, Davighi:2021oel, Chung:2021xhd}. Other frameworks include
  the 3-3-1 model with $\beta = -\sqrt{3}$ \cite{Gauld:2013qja,
    Buras:2013dea} and Branco-Grimus-Lavoura (BGL) models
  \cite{Celis:2015ara}.

\item $\CC_{9\mu}^{\NP} = -\CC_{10\mu}^{\NP}$. A variety of models
  have been proposed in which it is simply assumed that the $Z'$
  couples only to LH fermions (like the $W^\pm$ in the SM). These can
  be found in Refs.~\cite{Sierra:2015fma, Belanger:2015nma,
    Celis:2016ayl, Cline:2017lvv, Alonso:2017uky, Chiang:2017hlj,
    King:2017anf, Cline:2017ihf, Falkowski:2018dsl, Baek:2018aru,
    Allanach:2018lvl, Capdevila:2020rrl}.

\item $\CC_{9\mu}^{\NP}$ and $\CC_{10\mu}^{\NP}$. There are other
  models in which the $Z'$ couplings involve both $\CC_{9\mu}^{\NP}$
  and $\CC_{10\mu}^{\NP}$. The relative size of the two WCs is
  dictated by some underlying dynamics of the model, see
  Refs.~\cite{Buras:2013qja, Falkowski:2015zwa, Celis:2015eqs,
    Chiang:2016qov, Ko:2017lzd, Allanach:2019iiy, Altmannshofer:2019xda}.

\item Light $Z'$ bosons. In the above models, $M_{Z'} = O({\rm TeV})$
  is assumed. But this is not necessarily required. In
  Refs.~\cite{Cline:2017lvv, Sala:2017ihs, Bishara:2017pje, Darme:2020hpo, Darme:2021qzw}, scenarios
  are presented in which the $Z'$ has a mass of a few GeV. (The
  couplings are also reduced so as to reproduce the preferred values
  of $\CC_{9,10 \mu}^{\NP}$.)

\end{enumerate}

\subsubsection{LQ models}

There are five spin-0 and five spin-1 LQs, denoted
$\Delta$ and $V$ respectively, with couplings \cite{Alonso:2015sja}
\bea
{\cal L}_\Delta & = & ( y_{L u} {\bar L} u + y_{e Q}\, {\bar e} i \tau_2 Q ) \Delta_{-7/6}
+ y_{L d}\, {\bar L} d \Delta_{-1/6}
+ ( y_{L Q}\, {\bar L}^c i \tau_2 Q + y_{eu} \, {\bar e}^c u ) \Delta_{1/3} \nn\\
&& +~y_{ed}\, {\bar e}^c d \Delta_{4/3}
+ y'_{L Q}\, {\bar L}^c i \tau_2 {\vec \tau} Q \cdot {\vec \Delta}'_{1/3} + h.c. \nn\\
{\cal L}_V & = & (g_{L Q}\, {\bar L} \gamma_\mu Q + g_{ed}\, {\bar e} \gamma_\mu d) V^\mu_{-2/3}
+ g_{eu} \, {\bar e} \gamma_\mu u V^\mu_{-5/3}
+ g'_{L Q}\, {\bar L} \gamma_\mu {\vec \tau} Q \cdot {\vec V}^{\prime \mu}_{-2/3} \nn\\
&& +~(g_{L d}\, {\bar L} \gamma_\mu d^c + g_{e Q}\, {\bar e} \gamma_\mu Q^c) V^\mu_{-5/6}
+ + g_{L u} \, {\bar L} \gamma_\mu u^c V^\mu_{1/6} + h.c.
\label{LQlist}
\eea
$V^\mu_{-5/3}$ and $V^\mu_{1/6}$ do not couple to
down-type quarks, and so are not of interest to us.  The other LQs
transform as follows under $SU(3)_c \times SU(2)_L \times U(1)_Y$:
\bea
& R_2 \equiv \Delta_{-7/6} : ({\bar 3}, 2, -7/6) ~~,~~~~
{\tilde R}_2 \equiv \Delta_{-1/6} : ({\bar 3}, 2, -1/6) ~~,~~~~
S_1 \equiv \Delta_{1/3} : ({\bar 3}, 1, 1/3) ~, & \nn\\
& {\tilde S}_1 \equiv \Delta_{4/3} : ({\bar 3}, 1, 4/3) ~~,~~~~
S_3 \equiv {\vec \Delta}'_{1/3} : ({\bar 3}, 3, 1/3) ~, & \\
& U_1 \equiv V^\mu_{-2/3} : ({\bar 3}, 1, -2/3) ~~,~~~~
U_3 \equiv {\vec V}^{\prime \mu}_{-2/3} : ({\bar 3}, 3, -2/3) ~~,~~~~
V_2 \equiv V^\mu_{-5/6} : ({\bar 3}, 2, -5/6) ~. & \nn
\label{LQtypes}
\eea
Note that here the hypercharge is defined as $Y = Q_{em} - I_3$.
$R_2$, ${\tilde R}_2$, etc.\ are the names given to these LQs in
Ref.~\cite{Sakaki:2013bfa}. We adopt this nomenclature here.

All of these LQs were explored at different times as potential
explanations of the $\bsmumu$ anomalies. In Ref.~\cite{Alok:2017jgr},
fits were done including each LQ individually, and it was found that
only $S_3$, $U_1$ and $U_3$ provide good fits. (The $S_3$, $U_1$ and
$U_3$ LQs were originally examined in Refs.~\cite{Hiller:2014yaa,
  Gripaios:2014tna, Varzielas:2015iva, Sahoo:2015wya},
\cite{Hiller:2017bzc, Cline:2017aed} and \cite{Becirevic:2016oho},
respectively.) In the case of the $U_1$, the best fit has $g_{ed}^{\mu
  b} \simeq 0$, so that all three LQ solutions have $\CC_{9\mu}^{\NP} =
-\CC_{10\mu}^{\NP}$.

We therefore see that the $\CC_{9\mu}^{\NP} = -\CC_{10\mu}^{\NP}$
explanation can be generated by a $Z'$ or a LQ, while the
$\CC_{9\mu}^{\NP}$ solution is purely $Z'$.\footnote{{ Producing the $\CC_{9\mu}^{\NP}$ solution with LQs requires multiple LQs, whose couplings and masses are exceedingly fine-tuned.}}
Ref.~\cite{Alok:2017jgr}, it was shown that measurements of CP
violation in $B \to K^* \mu^+ \mu^-$ have the potential to distinguish
the $\CC_{9\mu}^{\NP}$ and $\CC_{9\mu}^{\NP} = -\CC_{10\mu}^{\NP}$
explanations. Similarly, it was argued in Ref.~\cite{Alok:2020bia}
that these two solutions can be differentiated through the
measurements of ${\cal B}(B_s \to \mu^+ \mu^-)$ and CP-averaged
azimuthal-angle asymmetries in $B \to K^* \mu^+ \mu^-$.

\section{\boldmath Charged-current anomalies: $\bclnu$}

We now turn to the charged-current $B$ flavour anomalies, observed in $\bclnu$ transitions. Whereas $\bsll$ occurs at loop level in the SM, $\bclnu$ is a tree-level decay in the SM (although the amplitude is multiplied by $V_{cb} \simeq 0.04$). Thus, a larger (tree-level) NP contribution is required to explain these anomalies. On the other hand, a similarity with the neutral-current anomalies is that both have been seen in LFUV observables.

\subsection{Experimental Results}

Discrepancies with the SM have been seen in the following observables:
\beq
R_{D^{(*)}} \equiv \frac{ {\cal B}({\bar B} \to D^{(*)} \tau^- {\bar\nu}_\tau) } { {\cal B}({\bar B} \to D^{(*)} \ell^- {\bar\nu}_\ell) } ~,~ \ell=e,\mu ~~,~~~~  
R_{J/\psi} \equiv \frac{ {\cal B}(B_c \to J/\psi \tau \nu_\tau)} { {\cal B}(B_c \to J/\psi \mu \nu_\mu)} ~.
\eeq
The latest results are (the $R_{D^{(*)}}$ numbers are taken from Ref.~\cite{HFLAV}, the $R_{J/\psi}$ numbers from Ref.~\cite{Watanabe:2017mip})
\bea
& R_D^{\tau/\ell} / (R_D^{\tau/\ell})_{\rm SM} = 1.14 \pm 0.10 ~~,~~~~
R_{D^*}^{\tau/\ell} / (R_{D^*}^{\tau/\ell})_{\rm SM} = 1.14 \pm 0.06 ~~, & \nn\\
& R_{J/\psi}^{\tau/\mu} / (R_{J/\psi}^{\tau/\mu})_{\rm SM} = 2.51 \pm 0.97 ~. &
\label{bctaunu_observ1}
\eea
The combined $R_D$ and $R_{D^*}$ results differ from the SM by $3.1\sigma$.

The above results suggest the presence of NP in $\bclnu$ decays, but it is not clear which of $\ell = e$, $\mu$ and/or $\tau$ is affected. However, it has also been found that
\beq
R_{D^*}^{e/\mu} / (R_{D^*}^{e/\mu})_{\rm SM} = 1.04 \pm 0.05 ~. 
\eeq
With this result, the simplest assumption is that NP  mainly affects $\bctaunu$, and most analyses focus on this scenario. With this in mind, two other relevant observables $B \to D^* \tau \nu_\tau$ are the $\tau$ polarization asymmetry $P_\tau(D^*)$ and the longitudinal $D^*$ polarization $F_L(D^*)$:
\bea
& P_\tau(D^*) \equiv \frac{ \displaystyle \Gamma( B \to D^* \tau^{\lambda=+1/2} \nu_\tau ) - \Gamma( B \to D^* \tau^{\lambda=-1/2} \nu_\tau )}
{\displaystyle \Gamma( B \to D^* \tau \nu_\tau ) } 
~, & \nn\\
& F_L(D^*) \equiv \frac{\displaystyle \Gamma( B \to D^*_L \tau \nu_\tau ) }{\displaystyle \Gamma( B \to D^* \tau \nu_\tau ) } ~. &
\label{bctaunu_observ2}
\eea
$P_\tau(D^*)$ and $F_L(D^*)$ were measured in Refs.~\cite{Belle:2016dyj, Belle:2017ilt}  and \cite{FLmeas}, respectively: 
\beq
P_\tau(D^*) = -0.38 \pm 0.51^{+0.21}_{-0.16}
~,~~
F_L(D^*) = 0.60 \pm 0.08 \pm 0.035 ~.
\eeq
These observables are useful for distinguishing NP explanation with different Lorentz structures. The unexpectedly large value observed for  $F_L(D^*)$ suggested to look for alternative ways to measure it (see \cite{Alguero:2020ukk}).

Finally, there is also the observable ${\cal B}(B_c \to \tau \nu_\tau)$, which has not yet been measured. In Ref.~\cite{Alonso:2016oyd} [Nov., 2016], it is argued that ${\cal B}(B_c \to \tau \nu_\tau) < 30\%$ is required for compatibility with the $B_c$ lifetime. This disfavours NP contributions to $\bctaunu$ with scalar couplings. In Ref.~\cite{Akeroyd:2017mhr} [Aug., 2017], it is claimed that LEP data requires ${\cal B}(B_c \to \tau^- {\bar\nu_\tau}) < 10\%$, placing an even stronger constraint on these NP scenarios. On the other hand, this question was re-examined in Ref.~\cite{Blanke:2018yud} [May, 2019], and it was concluded that the constraint ${\cal B}(B_c \to \tau^- {\bar\nu}_\tau) < 10\%$ is not justified. A conservative analysis yields ${\cal B}(B_c \to \tau^- {\bar\nu}_\tau) < 60\%$, removing the constraint on NP with scalar couplings. At the present time, the value of the allowed upper limit on ${\cal B}(B_c \to \tau^- {\bar\nu}_\tau)$ is an important open question. Of course, this could be resolved if this branching ratio were measured.

In the above, we have always assumed lepton flavour conservation (LFC), writing the leptons in the final state as $\tau^- {\bar\nu}_\tau$, $\mu^- {\bar\nu}_\mu$, etc. This is done principally because this is how the experimental results have been reported. However, it should be remembered that, if NP is present, LFC might not be followed. (On the other hand, in the SM, one {\it does} have LFC, so only NP couplings that obey LFC can interfere with the SM.)

\subsection{EFT analysis}

Because the ${\bar\nu}$ in $\bclnu$ decays is not detected, we know nothing about its properties. In particular, we don't know if it is LH or RH. The case of a RH final-state neutrino will be addressed in the model analysis below; EFT analyses assume LH neutrinos.

With this assumption, and assuming that it is $\bctaunu$ decays that receive NP contributions, the most general effective Hamiltonian for this decay is\footnote{As discussed above, here we assume lepton flavour conservation. For an EFT analysis of $\bclnu$ and $\bulnu$ decays without LFC, see Ref.~\cite{Jung:2018lfu}.}
\beq
H_{\rm eff} = \frac{4 G_F}{\sqrt{2}} \, V_{cb} \left[ (1 + C_V^L) O_V^L + C_V^R O_V^R + C_S^R O_S^R + C_S^L O_S^L + C_T O_T \right] ~,
\eeq
with
\bea
& O_V^L = ( {\bar c} \gamma^\mu P_L b ) ( {\bar \tau} \gamma_\mu P_L \nu_\tau ) ~~,~~~~ 
O_V^R = ( {\bar c} \gamma^\mu P_R b ) ( {\bar \tau} \gamma_\mu P_L \nu_\tau ) ~, & \nn\\
& O_S^R = ( {\bar c} P_R b ) ( {\bar \tau} P_L \nu_\tau ) ~,~~ 
O_S^L = ( {\bar c} P_L b ) ( {\bar \tau} P_L \nu_\tau ) ~,~~ 
O_T = ( {\bar c} \sigma^{\mu\nu} P_L b ) ( {\bar \tau} \sigma_{\mu\nu} P_L \nu_\tau ) ~.
\label{bctaunuops}
\eea
Here the SM contribution has been factored out, so that $C_{V,S,T}^{L,R}$ are generated only through NP. 

As was the case in the $\bsll$ EFT analysis, these operators are
defined in the WET. In order to make contact with the NP that
generates them, they must be mapped onto SMEFT. It turns out that all
of these dimension-6 WET operators are also dimension-6 SMEFT
operators, with one exception: $O_V^R$. An LFU version of this
operator appears in SMEFT at dimension-6, but the LFUV operator shown
above arises only at dimension 8 \cite{Burgess:2021ylu}, so that its
coefficient $C_V^R$ is expected to be smaller than the other
coefficients by $O(v^2/\Lambda^2)$, where $v$ is the vev of the SM
Higgs and $\Lambda$ is the scale of NP. For this reason, $O_V^R$ is
usually excluded from EFT analyses. (For an alternative point of view,
see Refs.~\cite{nonSMEFTNP, Cata:2015lta}.)

The most recent EFT analysis of the $\bctaunu$ anomalies was performed in Ref.~\cite{Blanke:2019qrx}. This study used the five observables of Eqs.~(\ref{bctaunu_observ1}) and (\ref{bctaunu_observ2}), and added ${\cal B}(B_c \to \tau \nu_\tau)$, assuming upper limits of 60\%, 30\% and 10\%. The results are shown in Table \ref{bctaunufitresults}. Note that tensor operators are generally generated by Fierz transformations. For example, the exchange of the scalar $SU(2)_L$-doublet LQ $R_2$ produces the four-fermion operator $( {\bar c} P_L \nu_\tau ) ( {\bar \tau} P_L b)$ \cite{Becirevic:2016yqi, Becirevic:2018afm}. A Fierz transformation produces both $O_S^L$ and $O_T$, with $C_S^L = 4 C_T$. This is used in the fits.

\begin{table}[h]
\begin{center}
\begin{tabular}{|c||c|c|c|}
  \hline
  1D hypothesis & Best fit & $p$ value (\%) & Pull$_{\rm SM}$ \\ 
\hline
\hline
$C_V^L$ & $0.07 \pm 0.02$ & 44 & 4.0 \\
\hline
$C_S^R$ & $0.09 \pm 0.03$ & 2.7 & 3.1 \\
\hline
$C_S^L$ & $0.07 \pm 0.03$ & 0.26 & 2.1 \\
\hline
$C_S^L = 4 C_T$ & $-0.03 \pm 0.04$ & 0.04 & 0.7 \\
\hline
\end{tabular}
\begin{tabular}{|c||c|c|c|}
  \hline
  2D hypothesis & Best fit & $p$ value (\%) & Pull$_{\rm SM}$ \\ 
\hline
\hline
$(C_V^L, C_S^L = 4 C_T)$ & $(0.10, -0.04)$ & 29.8 & 3.6 \\
\hline
$\left. (C_S^R, C_S^L)\right\vert_{60\%}$
& $\displaystyle{(0.29, -0.25) \atop (-0.16, -0.69)} $ & 75.7 & 3.9 \\
\hline
$\left. (C_S^R, C_S^L)\right\vert_{30\%}$
& $\displaystyle{(0.21, -0.15) \atop (-0.26, -0.61)} $ & 30.9 & 3.6 \\
\hline
$\left. (C_S^R, C_S^L)\right\vert_{10\%}$
& $\displaystyle{(0.11, -0.04) \atop (-0.37, -0.51)} $ & 2.6 & 2.9 \\
\hline
$(C_V^L, C_S^R)$ & $(0.08, 0.01)$ & 26.6 & 3.6 \\
\hline
$\left. ({\rm Re}[C_S^L = 4 C_T],{\rm Im}[C_S^L = 4 C_T])\right\vert_{60,30\%}$
& $(-0.06, \pm 0.31)$ & 25.0 & 3.6 \\
\hline
$\left. ({\rm Re}[C_S^L = 4 C_T],{\rm Im}[C_S^L = 4 C_T])\right\vert_{10\%}$
& $(-0.03, \pm 0.24)$ & 5.9 & 3.2 \\
\hline
\end{tabular}
\end{center}
\caption{Results of the one-dimensional and two-dimensional fits for
  the WCs (given at the matching scale of 1 TeV), including the five
  observables of Eqs.~(\ref{bctaunu_observ1}) and
  (\ref{bctaunu_observ2}), with ${\cal B}(B_c \to \tau \nu_\tau) <
  60\%$, 30\% and 10\%.  For those entries without a label indicating
  the constraint on ${\cal B}(B_c \to \tau \nu_\tau)$, the fit is
  valid for all three upper limits.  For the SM, the $p$-value is $\sim 0.1\%$, which represents a discrepancy with the data of $3.3\sigma$.}
\label{bctaunufitresults}
\end{table}

From this Table, we make the following observations:
\begin{itemize}

\item Among the 1D hypotheses, only a nonzero $C_V^L$ provides a good
  explanation of the data.

\item There are four 2D hypotheses that are good fits. Two of them --
  $(C_V^L, C_S^L = 4 C_T)$ and $(C_V^L, C_S^R)$ -- are basically just
  the 1D best fit, with a small additional WC added. The
  goodness-of-fit of the other two -- $(C_S^R, C_S^L)$ and $({\rm
    Re}[C_S^L = 4 C_T],{\rm Im}[C_S^L = 4 C_T])$ -- depends critically
  on the upper limit of ${\cal B}(B_c \to \tau \nu_\tau)$. If this
  upper limit is 30\% or 60\%, these 2D hypotheses are
  viable. However, if the upper limit is 10\%, these scenarios are
  disfavoured.

\item We note that the difference between $p$ values of 30\% and 75\%
  is not statistically significant -- both describe perfectly
  acceptable fits. In fact, from a statistical point of view, we
  expect the true solution to have a $p$ value of $\sim 50\%$. Larger
  $p$ values describe fits that are ``too good,'' and suggest that
  errors may be overestimated. In this sense, it may be that allowing
  ${\cal B}(B_c \to \tau \nu_\tau)$ to be as large as 60\% is a bit
  too weak of a constraint. (Of course, this discussion will become
  moot with an actual measurement of ${\cal B}(B_c \to \tau
  \nu_\tau)$.)

\end{itemize}

\subsection{Model analysis}
\label{bctaunu_models}

When the $R_{D^{(*)}}$ anomalies were first announced, there were several different general EFT analyses that identified the WCs that could be modified in order to explain the data \cite{Fajfer:2012jt, Tanaka:2012nw, Freytsis:2015qca, Alok:2017qsi}. Although they were model-independent, they did indicate what types of NP models should be considered. The key point is that $\bctaunu$ is a charged-current process. Given that it is produced at tree level in the SM, the simplest models involve the tree-level exchange of a $W'$, a LQ, or a charged Higgs. We discuss these possibilities in turn, often within the context of the above EFT results.

\subsubsection{\boldmath $W'$}

For the $\bctaunu$ anomaly, the $W'$ solution most often proposed involves a RH $W_R$, along with a light RH neutrino \cite{Asadi:2018wea, Greljo:2018ogz, Robinson:2018gza, Babu:2018vrl}. A recent complete analysis can be found in Ref.~\cite{Mandal:2020htr}. A generic $W'$ model, with arbitrary LH and RH couplings to both quarks and leptons, is examined in Ref.~\cite{Gomez:2019xfw}.

\subsubsection{LQ models}

The first paper to examine LQ models as a possible explanation of the $\bctaunu$ anomaly is Ref.~\cite{Sakaki:2013bfa}. Followup papers can be found in Refs.~\cite{Li:2016vvp, Bansal:2018nwp, Aydemir:2019ynb, Cheung:2020sbq}. At present, the $S_1$, $U_1$ and $R_2$ LQs [see Eq.~(\ref{LQtypes})] can be viable explanations. The exchange of an $S_1$ or $U_1$ leads to a four-fermion operator that is equivalent to $O_V^L$ of Eq.~(\ref{bctaunuops}). As for $R_2$, as mentioned above, its exchange leads to $O_S^L$ and $O_T$, with $C_S^L = 4 C_T$. So it can provide an explanation only for large values of ${\cal B}(B_c \to \tau \nu_\tau)$.

\subsubsection{Charged Higgs}

Another way of explaining the $\bctaunu$ anomaly is through the
tree-level exchange of a charged Higgs boson, and here numerous
two-Higgs-doublet models (2HDMs) have been proposed
\cite{Crivellin:2012ye, Celis:2012dk, Crivellin:2015hha, Wei:2017ago,
  Chen:2017eby, Lee:2017kbi, Iguro:2017ysu, Martinez:2018ynq,
  Fraser:2018aqj, Cardozo:2020uol}. However, charged-Higgs exchange
produces only $O_S^R$ and $O_S^L$ of Eq.~(\ref{bctaunuops}), and as we
saw in the EFT analysis, this solution is only viable if ${\cal B}(B_c
\to \tau \nu_\tau)$ is large. In addition, there are strong
constraints on the charged Higgs from the LHC \cite{Iguro:2018fni}.

\bigskip

A variety of measurements have been proposed to distinguish among the NP explanations. These include the $q^2$ distribution \cite{Sakaki:2014sea}, $D^*$ polarization \cite{Alok:2016qyh, Iguro:2018vqb}, $\tau$ polarizations \cite{Ivanov:2017mrj}, and the full angular distribution \cite{Bhattacharya:2020lfm}.

\section{\boldmath Links between the $\bsmumu$ and $\bctaunu$ anomalies}

\subsection{SMEFT analysis}

It was first shown in Ref.~\cite{Bhattacharya:2014wla} that it is possible to simultaneously explain the $\bsmumu$ and $\bctaunu$ anomalies. At that time (2014), it was thought that NP with purely LH couplings was required to account for the $\bsmumu$ data \cite{Hiller:2014yaa}. Referring to Eq.~(\ref{SMEFTops}), we see that there are two SMEFT operators of this type:
\beq
{\cal O}_{ijkl}^{(1)} = ({\bar L}_i \gamma_\mu  L_j) ({\bar Q}_k \gamma^\mu Q_l) ~~,~~~~
{\cal O}_{ijkl}^{(3)} = ({\bar L}_i \gamma_\mu \sigma^I  L_j) ({\bar Q}_k \gamma^\mu \sigma^I Q_l) ~, 
\eeq
where we have included the flavour indices in the names of the operators. 

The key point is that, while ${\cal O}_{ijkl}^{(1)}$ is purely neutral
current, ${\cal O}_{ijkl}^{(3)}$ contains both neutral-current and
charged-current contributions. Thus, ${\cal O}_{2223}^{(1,3)}$
generate $\bsmumu$, ${\cal O}_{3323}^{(3)}$ produces $\bctaunu$. The
original idea of Ref.~\cite{Bhattacharya:2014wla} (see also
Refs.~\cite{Greljo:2015mma, Calibbi:2015kma}) was that the NP couples
only (or mainly \cite{Greljo:2015mma}) to the third generation in the
gauge basis. Couplings to the second generation are generated when one
rotates to the mass basis. Since ${\cal O}_{2223}^{(1,3)}$ requires
more such rotations than ${\cal O}_{3323}^{(3)}$, ${\cal
  C}_{2223}^{(1,3)}$ is naturally smaller than ${\cal
  C}_{3323}^{(3)}$. This is what is required since $\bsmumu$ is loop
level in the SM, while $\bctaunu$ is tree level. In
Ref.~\cite{Kumar:2018kmr}, it was shown that, even if one makes no
assumptions about the coupling of the NP in the gauge basis, when all
constraints are taken into account, one finds the same pattern as
above for the WCs. Thus, both $B$ anomalies can be explained
simultaneously if the NP is purely LH and generates ${\cal
  O}_{ijkl}^{(1)}$ and ${\cal O}_{ijkl}^{(3)}$.

But there's more. In addition to producing $\bctaunu$, ${\cal O}_{3323}^{(3)}$ also generates $\bstautau$ \cite{Capdevila:2017iqn}. This has two effects. First, it implies that the branching ratios of decays governed by $\bstautau$ should be much larger than what is predicted by the SM. Second, this coupling is precisely what is required to produce $\CC_{9}^U$ in $\bsmumu$ at one loop (see Sec.~\ref{LFUVLFU}). 

Thus, this combined explanation of the two $B$ anomalies reproduces the favoured $[\CC_{9}^V=-\CC_{10}^V,\CC_{9}^U]$ solution (see Eq.~(\ref{LFUVLFUsoln})) of the $\bsmumu$ anomalies. The combined Pull$_{\rm SM}$ of this particular scenario is as high as 8.1$\sigma$. The central values found for the WCs in this case are $\Cc{9}^V=-\Cc{10}^V=-0.36, \Cc{9}^U=-0.74$ \cite{Alguero:2021anc}.

\subsection{Model analysis}

As shown above, it is possible to find a combined explanation of the $\bsmumu$ and $\bctaunu$ anomalies if the NP generates the operators ${\cal O}_{ijkl}^{(1)}$ and ${\cal O}_{ijkl}^{(3)}$. That is, the NP must couple to LH particles and have the appropriate couplings to produce $\bsmumu$ and $\bctaunu$.  

Now, in an EFT analysis, one can focus on a particular operator. However, in a model, additional operators will generally be produced, and these lead to additional constraints. Thus, in order to test whether a particular model is viable, one must consider all of these constraints. 

\subsubsection{\boldmath VB ($W', Z'$) models}

The vector-boson (VB) model, which contains a SM-like $SU(2)_L$ triplet of vector bosons ($W', Z'$), can potentially provide a simultaneous explanation of both $B$ anomalies. The $Z'$ and $W'$ respectively mediate the $\bsmumu$ and $\bctaunu$ transitions, and their couplings can be chosen to reproduce the experimental data. Explicit VB models have been proposed in Refs.~\cite{Greljo:2015mma, Boucenna:2016qad}.

On the other hand, there are additional constraints on the VB model. These include the constraints on the $Z'$ couplings described in Sec.~\ref{Z'models}, but there are also constraints from lepton-flavour-violating processes. The VB model has been examined in detail in Refs.~\cite{Kumar:2018kmr, Bhattacharya:2016mcc, Buttazzo:2017ixm}. There it is found that, when all the constraints are taken into account, the VB model is excluded as a simultaneous explanation of both anomalies.

\subsubsection{LQ models}
\begin{enumerate}

\item Single-LQ solutions: Over the years, many single-LQ solutions
  were proposed as a simultaneous solution of the $\bsmumu$ and
  $\bctaunu$ anomalies. As new data appeared, certain solutions were
  ruled out.  The combined EFT analysis in
  Ref.~\cite{Buttazzo:2017ixm} clearly demonstrated that the only
  viable single-LQ solution is the one based on the vector LQ $U_1$,
  which was first discussed in Refs.~\cite{Alonso:2014csa,
    Calibbi:2015kma, Barbieri:2015yvd}. This conclusion was further
  confirmed in Ref.~\cite{Angelescu:2021lln}, which considered all
  single-LQ solutions, taking into account LHC constraints.

\item Two-LQ solutions: In Ref.~\cite{Crivellin:2017zlb}, it is
  proposed to simultaneously explain the anomalies with the addition
  of two scalar LQs, $S_1$ and $S_3$. $S_3$ contributes to both
  $\bsmumu$ and $\bctaunu$, but it cannot explain both anomalies by
  itself. $S_1$ contributes to $\bctaunu$, but not $\bsmumu$. And both
  LQs contribute to $\bsnunubar$. It is possible to choose the
  couplings of $S_1$ and $S_3$ such that both anomalies are explained,
  while evading the constraints from $\bsnunubar$.

\end{enumerate}

\subsubsection{UV Completions}

The idea of explaining both anomalies with a single NP particle is
very intriguing, so the $U_1$ solution has generated a great deal of
interest. One problem with this solution is that, since the $U_1$ is a
vector LQ, the theory with a $U_1$, but without a Higgs sector, is not
renormalizable. As a consequence, loop diagrams, which can lead to
potentially important effects, cannot be calculated. In order to
address this problem, a number of papers have constructed models with
an extended gauge group that is broken at the TeV scale and contains
the $U_1$ LQ as a gauge boson, or as a composite state. A key
observation, made first in Ref.~\cite{Barbieri:2015yvd}, is that the
$U_1$ field points to a unification of quarks and leptons via a local
or global $SU(4)$ symmetry.

These UV completion models can be split into two categories. First,
there are the models that use variations of the Pati-Salam idea, in
which $SU(4)_{PS}$ unifies $SU(3)_C$ and a $U(1)$ under which both
quarks and leptons are charged (e.g., $U(1)_Y$ or $U(1)_{B-L}$)
\cite{Assad:2017iib, DiLuzio:2017vat, Calibbi:2017qbu,
  Bordone:2017bld, Barbieri:2017tuq, Blanke:2018sro, Aydemir:2018cbb,
  Heeck:2018ntp, Balaji:2018zna, Fornal:2018dqn,
  Balaji:2019kwe, Iguro:2021kdw}. Second, some models use the ``4321'' gauge group,
$SU(4) \times SU(3)' \times SU(2)_L \times U(1)_X$
\cite{Greljo:2018tuh, Cornella:2019hct, Guadagnoli:2020tlx}. In all
cases, it must be ensured that the breaking of the gauge group at the
TeV scale does not run into any problems from low-energy flavour
constraints.

\section{Connections with other physics problems}

There are several other experimental results that cannot be explained by the SM. If NP is present, its effects could possibly be seen in different types of observables. With this in mind, models have been proposed that simultaneously explain the $B$ anomalies and (at least) one of these other discrepancies with the SM. { Below we discuss models that relate the $B$ anomalies (mostly $\bsll$) to dark matter, $(g-2)_\mu$, neutrino masses, and hadronic anomalies. But there are also attempts to make connections with other observables that exhibit tensions with the SM, such as the Cabibbo angle anomaly, the $Z\to \bar{b}b$ forward-backward asymmetry, and $\tau \to \mu {\bar\nu}\nu$, see Ref.~\cite{Crivellin:2020oup} for example.} 

\subsection{Dark matter}

There is considerable cosmological and astrophysical evidence
supporting the existence of dark matter (DM). However, the fact that
the SM has no DM candidate is another one of its shortcomings, and
provides another reason for the requirement of NP. A number of papers
have explored possible links between these two strong hints of NP, the
$\bsll$ anomalies and DM. In Ref.~\cite{Vicente:2018frk}, these models
are separated into two categories: (i) portal models, in which the
relic density mediator is the same as that in the $B$ anomalies, and
(ii) loop models, in which the $\bsll$ anomalies are explained via
loops that include DM. In the discussion of $\bsll$ models in
Sec.~\ref{bsllmodels}, we elected to focus only on the simplest
models. These include only models of type (i); models of type (ii)
contain many more particles, are subject to more constraints, and are
often valid only in a corner of parameter space. Portal models with a
$Z'$ mediator can be found in Refs.~\cite{Sierra:2015fma,
  Belanger:2015nma, Celis:2016ayl, Altmannshofer:2016jzy, Ko:2017quv,
  Ko:2017yrd, Cline:2017lvv, Falkowski:2018dsl, Arcadi:2018tly,
  Hutauruk:2019crc, Biswas:2019twf, Han:2019diw,
  Borah:2020swo}. (These do not include models in which the $Z'$ is
allowed to decay to DM in order to evade LHC constraints, but do not
otherwise address the DM issues of relic density and direct
detection.) Refs.~\cite{Cline:2017aed, Choi:2018stw,
  Guadagnoli:2020tlx, Baker:2021llj}, which include combined
explanations of both the $\bsll$ and $\bclnu$ anomalies, describe
portal models with LQs.

\subsection{\boldmath $(g-2)_\mu$}

Recently, the Muon $g-2$ Collaboration released a new result: the
measurement of $a_\mu \equiv (g-2)_\mu/2$ disagrees with the SM
prediction by $4.2\sigma$ \cite{Muong-2:2021ojo}. Following this
announcement, a number of papers were written proposing models to
simultaneously explain the observed value of $a_\mu$ and one or both
of the $B$ anomalies. There are many different scenarios, but all of
them use a $Z'$ and/or LQs, see Refs.~\cite{Greljo:2021xmg,
  Navarro:2021sfb, Wang:2021uqz, Greljo:2021npi, Perez:2021ddi,
  Ban:2021tos, Du:2021zkq, Marzocca:2021azj, Cen:2021iwv, Darme:2020hpo, Darme:2021qzw}.

\subsection{Neutrino properties}

Another weakness of the SM is that it has no explanation for neutrino
masses. Here too, models have been proposed that provide combined
explanations of the $B$ anomalies and neutrino masses. However, it
must be said that the connection between the two is usually rather
tenuous in these models, and they generally involve several different
types of NP particles. See Refs.~\cite{Dev:2020qet, Saad:2020ihm, Babu:2020hun,
  Bigaran:2019bqv} for recent examples of such models.
  
\subsection{Hadronic anomalies}

The $\bsll$ and $\bclnu$ anomalies both involve semileptonic transitions. However, there is also some tension with the SM in certain purely hadronic $B$ decays, albeit at a much lower level. Attempts have been made to link the semileptonic and hadronic anomalies.

For many years, the measurements of the branching ratios and direct \& indirect CP asymmetries in $B \to \pi K$ decays ($B^0 \to \pi^- K^+$, $B^0 \to \pi^0 K^0$, $B^+ \to \pi^0 K^+$, $B^+ \to \pi^+ K^0$) have exhibited a slight inconsistency. This can be ameliorated if there is a NP contribution to the $b \to s$ electroweak penguin. In Ref.~\cite{Beaudry:2017gtw}, an update of the ``$B \to \pi K$ puzzle'' was performed, and it was noted that this could be explained if the $Z'$ used to explain the $\bsmumu$ anomaly also couples to RH $u{\bar u}$ and/or $d{\bar d}$, with unequal couplings.

U spin is an approximate symmetry that treats $d$ and $s$ quarks as identical. Under U spin, the two transitions $b \to s$ and $b \to d$ are equal, apart from CKM matrix elements. Note that these two amplitudes involve second-generation and first-generation {\it quarks} in the final state. A certain parallel may be drawn with $b \to s \mu^+\mu^-$ and $b \to s e^+e^-$, which involve second-generation and first-generation {\it leptons} in the final state. In the latter case, the ratio of the two amplitudes tests LFUV (involving QED), while in the former it measures U-spin breaking (involving QCD). This shows the limitations of the analogy. 


Inspired by this observation, the authors of Ref.~\cite{Alguero:2020xca} constructed an optimized observable, $L_{K^* \bar{K}^*}$, from the ratio of longitudinal amplitudes for $B_s \to {\bar K}^{0*} K^{0*}$ ($b \to s$) and $B_d \to {\bar K}^{0*} K^{0*}$ ($b \to d$) that can be predicted (in contrast to the other amplitudes in which infrared-divergences enter at leading order). The results were as follows:
\begin{itemize}

    \item  The U-spin breaking in the ratio was computed within the SM using QCDF and modeling power-suppressed but infrared-divergent weak annihilation and hard-spectator scattering. Taking this into account, there is a tension with the SM at the level of  2.6$\sigma$ in $L_{K^* \bar{K}^*}$.
    
    \item The $b \to s$ amplitude is smaller than the $b \to d$ amplitude in the same way as there is a deficit in $b \to s \mu^+\mu^-$ as compared to  $b \to s e^+e^-$.
    
    \item There is a dominant operator that can naturally explain the low value of 
    $L_{K^*\bar{K}^*}$ as it happens for $b\to s\ell\ell$ with $O_9$. The Wilson coefficient of this dominant operator also
    gets a  destructive NP contribution with a $\sim 20\%$  of the SM size. 
    
    \item A possible connection with the $b \to s\ell\ell$ anomalies is to consider a composite/extra-dimensional model whose particle spectrum includes a Kaluza-Klein gluon and a $Z^\prime$.
    
\end{itemize}



\section{An EFT glance into the future}

The future prospects for the field of $B$ flavour anomalies are very exciting -- we expect a plethora of new data and updates of previous measurements. Indeed, in the coming five years, the search for NP using these anomalies is expected to play a leading role at CERN and particularly at Belle-II.

\subsection{\boldmath $\bsmumu$ anomalies}

In the very near future, we expect to receive two different types of input:
\begin{itemize}

\item LFUV observables. In the short term, measurements of $R_{K_S}$, $R_{K^{*+}}$ (based on the charged channel $B^+ \to K^{*+}\ell^+\ell^-$) and $R_\phi$ (based upon $B_s \to\phi\ell\ell$) are expected from LHCb. These results will test whether or not there is LFUV in Nature, which is what the present  measurements of $R_{K^*}$ and especially $R_K$ are clearly pointing to. In the medium term, once more data is accumulated, an update of the observable $R_K$, possibly increasing its significance much beyond the present evidence for NP, will be a crucial milestone.

\item Angular optimized observables. Here we can expect updates of the
  optimized observables of the 4-body angular distributions of $B_d^0
  \to K^* (\to K\pi) \, \mu^+ \mu^-$ (the last update was in 2020
  \cite{LHCb:2020lmf}), the corresponding $B^+$ decay, and $B_s^0 \to
  \phi (\to K^+ K^-) \, \mu^+ \mu^- $. All of these will be important,
  but the first one will be particularly interesting, to see if the
  first anomaly, that is the observable $P_5^\prime$, increases with
  more data.

In addition, an important next step in this program will be the study of the angular observables entering the $S$-wave part of the $B \to K\pi \ell^+\ell^-$ decay. These new $S$-wave observables will provide complementary information to the $P$-wave part of the distribution \cite{Alguero:2021yus}. 

    
\end{itemize}

Finally, at present, the two scenarios with the largest Pull$_{\rm SM}$ are the 2D hypotheses $[\Cc{9\mu}^{\rm NP} , \Cc{9\mu}^{\rm \prime NP} = -\Cc{10\mu}^{\rm \prime NP}]$ and $[{\cal C}^V_{9}=-{\cal C}^V_{10}, {\cal C}^U_9]$ (Eq.~(\ref{finalscenarios})). It will be interesting to see which of these dominates as the preferred explanation, or if another solution emerges. It was shown in Refs.~\cite{Alguero:2019ptt,Alguero:2021anc, Alguero:2019pjc} that, while $R_K$ and $R_{K^*}$ cannot disentangle between these two scenarios, a precise measurement of the observable $Q_5$ in the bin [1.1,6] GeV$^2$ can separate them, with $Q_5 \sim 0.08$ for the second scenario, and $Q_5 \sim 0.32$ for the first.

\subsection{\boldmath $\bctaunu$ anomalies}

At the end of Sec.~\ref{bctaunu_models}, we note that a variety of measurements have been proposed to distinguish among the NP explanations. There is one that is particularly promising: it can, in principle, pinpoint what type of NP is present. Ref.~\cite{Bhattacharya:2020lfm} proposes measuring the angular distribution in ${\bar B} \to D^* (\to D \pi') \, \tau^{-} (\to \pi^- \nu_\tau) {\bar\nu}_\tau$. By measuring the four-momenta of the final-state $D$, $\pi'$ and $\pi^-$, the differential decay rate can be constructed. This depends on two non-angular variables, $q^2$ and $E_\pi$, and three angles. Here, $q^2$ is the invariant mass-squared of the $\tau^-{\bar\nu}_\tau$ pair and $E_\pi$ is the energy of the $\pi^-$ in the $\tau$ decay. The idea is then to separate the data into various $q^2$-$E_\pi$ bins, and then to perform an angular analysis within each bin. The angular distribution consists of twelve different angular functions; nine of these are CP-conserving, and three are CP-violating. We therefore potentially have a large number of observables in this differential decay rate; the exact number depends on how many $q^2$-$E_\pi$ bins there are.

In the angular distributions, the coefficients of each angular function are functions of $q^2$, $E_\pi$, and the NP coefficients of Eq.~(\ref{bctaunuops}). There are five NP operators; only one linear combination of the two scalar operators contributes to the decay ($O_S^L + O_S^R$ does not).  With complex coefficients, there are eight unknown theoretical parameters. Thus, if the angular distribution in ${\bar B} \to D^* (\to D \pi') \, \tau^{-} (\to \pi^- \nu_\tau) {\bar\nu}_\tau$ can be measured, it may be possible to extract all of the NP coefficients from a fit, thus revealing what type of NP is present.

\section{Summary}
\label{Conclusions}

{ Despite its enormous success, the SM of
  particle physics is not complete -- there must exist physics beyond
  the SM. Direct searches at the LHC for this new physics have thus
  far proven fruitless. On the other hand, there are deviations from
  the predictions of the SM in a number of observables involving
  $\bsll$ and $\bclnu$ decays. These are the $B$ flavour anomalies,
  and they provide an indirect hint of NP.

The $\bsll$ anomalies have been seen in a number of different
observables, measured by different experimental groups. At present,
the p-value of the SM fit to all the $b \to s \ell\ell$ data is $\sim 1.1\%$,  corresponding to a discrepancy with the data of $\sim 2.5\sigma$. Global fits to
the data have been performed by different groups. Despite having
different methodologies, the results are remarkably consistent, particularly between the ACDMN and HMMN} groups. Using
the results of the ACDMN group \cite{Alguero:2021anc}, the preferred
1D solutions are $\Cc{9\mu}^{\rm NP}$ (Pull$_{\rm SM} = 7.0\sigma$) and
$\Cc{9\mu}^{\rm NP}=-\Cc{10\mu}^{\rm NP}$ (6.2$\sigma$), and the preferred 2D
solutions are $[\Cc{9\mu}^{\rm NP} , \Cc{9\mu}^{\rm \prime NP} =
  -\Cc{10\mu}^{\rm \prime NP}]$ (7.4$\sigma$) and $[{\cal C}^V_{9}=-{\cal
    C}^V_{10}, {\cal C}^U_9]$ (7.3$\sigma$). The simplest models involve the
tree-level exchange of a $Z'$ or a LQ.

{
The $\bclnu$ anomalies have been seen in fewer observables than
$\bsll$, but here too, they have been measured by different
experiments, with the results largely confirmed. { The $p$-value of the
SM fit to the present data is $\sim 0.1\%$, corresponding to a
discrepancy of $3.3\sigma$.} The question of the
identification of the underlying NP is complicated by the fact that we do
not know if the neutrino is LH or RH. Assuming a LH neutrino, the
preferred solution in the global fit is $C_V^L$ i.e., LH NP,  with
Pull$_{\rm SM} = 4.0\sigma$. Models involve the tree-level exchange of a
$W'$ or a LQ. If ${\cal B}(B_c \to \tau \nu_\tau)$ is allowed to be as
large as 60\%, $(C_S^R, C_S^L)$, i.e., scalar NP such as a charged
Higgs, also provides a good solution. 

It is obviously far too early to bet on a particular NP explanation,
but one intriguing possibility is the vector LQ $U_1$. It can
simultaneously explain both anomalies, and in the case of $\bsll$, it
can generate one of the preferred solutions, $[{\cal C}^V_{9}=-{\cal
    C}^V_{10}, {\cal C}^U_9]$.

These are exciting times. We may be seeing indirect evidence of NP at
low energies. In the near future, LHCb and especially Belle II will
produce new results.  These will include measurements of
previously-measured quantities, as well as new ones, such as $Q_5$,
observables involving $b \to s \tau^+\tau^-$, and ${\cal B}(B_c \to
\tau \nu_\tau)$. All of this information will give us a better idea of
what the underlying NP is.


\bigskip
\bigskip
{\it Note added:} As this review was being finalized, the measurements of $R_{K_S}$ and $R_{K^{*+}}$ were reported by LHCb \cite{LHCb:2021lvy}. $R_{K_S}$ was measured in the bin [1.1,6]~GeV$^2$, while $R_{K^{*+}}$ was measured in a wider bin, [0.045,6.0]~GeV$^2$, and exhibit tensions with the SM of 1.5$\sigma$ and 1.4$\sigma$, respectively. These are not large disagreements, but they are in the same direction as previous measurements, and reinforce the case for LFUV. 

\bigskip

{\bf Disclosure Statement}: The authors are not aware of any
affiliations, memberships, funding, or financial holdings that might
be perceived as affecting the objectivity of this review.

\bigskip

{\bf Acknowledgments}: We would like to acknowledge useful discussions
with Andreas Crivellin, Gino Isidori, Avelino Vicente and Nazila
Mahmoudi. The work of DL was partially financially supported by NSERC
of Canada.  This work received financial support from the Spanish
Ministry of Science, Innovation and Universities (FPA2017-86989-P) and
the Research Grant Agency of the Government of Catalonia (SGR 1069)
[JM].  JM acknowledges the financial support by ICREA under the ICREA
Academia programme.
%




\bibliographystyle{<ar-style3.bst>}

\end{document}